\documentclass[epj]{svjour}

\usepackage{color}
\usepackage{xcolor}
\definecolor{bred}{rgb}{0.8, 0.0, 0.0}
\definecolor{pblue}{rgb}{0.2, 0.2, 0.6}
\definecolor{ao}{rgb}{0.0, 0.5, 0.0}
\definecolor{carmine}{rgb}{0.59, 0.0, 0.09}

\newcommand{\keVee}{\,keV$_{ee}$}
\newcommand{\keVnr}{\,keV$_{nr}$}
\newcommand{\degree}{$^{\text{o}}$}
\newcommand{\CEvNS}{CE$\nu$NS}

\usepackage{graphicx}
\usepackage{amsmath}
\usepackage{amssymb}
\usepackage{hyperref}
\usepackage{cite}
\usepackage{arydshln}
\usepackage{upgreek}
\usepackage{subcaption}

\usepackage{hyperref}
\hypersetup{
    colorlinks = true,
    linkbordercolor = {white},
    linkcolor = bred,
    citecolor = pblue,
    urlcolor = pblue
}

\usepackage[switch]{lineno} 

\begin{document}\sloppy

\title{Direct measurement of the ionization quenching factor of nuclear recoils in germanium in the keV energy range}

\titlerunning{Direct measurement of the ionization quenching factor of nuclear recoils in Ge in the keV range}
\author{A.~Bonhomme\inst{1}\thanks{\emph{Corresponding author:} aurelie.bonhomme@mpi-hd.mpg.de, conus.eb@mpi-hd.mpg.de}, H.~Bonet\inst{1}, C.~Buck\inst{1}, J. Hakenm\"{u}ller\inst{1}, G.~Heusser\inst{1}, T.~Hugle\inst{1}, M.~Lindner\inst{1}, W.~Maneschg\inst{1}, R.~Nolte\inst{2}, T.~Rink\inst{1}, E.~Pirovano\inst{2}, H.~Strecker\inst{1}
}
\authorrunning{A. Bonhomme et al.}
\institute{Max-Planck-Institut f\"ur Kernphysik, Saupfercheckweg 1, 69117 Heidelberg, Germany \and Physikalisch-Technische Bundesanstalt, Bundesallee 100, 38116 Braunschweig, Germany}

\date{}

\abstract{
		This article reports the measurement of the ionization quenching factor in germanium for nuclear recoil energies in the keV range.
		Precise knowledge of this factor in this energy range is highly relevant for coherent elastic neutrino-nucleus scattering and low mass dark matter searches with germanium-based detectors.
		Nuclear recoils were produced in a thin high-purity germanium target with a very low energy threshold via irradiation using monoenergetic neutron beams. 
		The energy dependence of the ionization quenching factor was directly measured via kinematically constrained coincidences with surrounding liquid scintillator based neutron detectors. The systematic uncertainties of the measurements are discussed in detail. With measured quenching factors between 0.16 and 0.23 in the 0.4\keVnr~to 6.3\keVnr~energy range, the data are compatible with the Lindhard theory with a parameter $k$ of 0.162\,$\pm\,0.004$ (stat+sys).
		}


\maketitle

\section{Introduction} 

Recent advances in the detection of very low energy signals opened up new possibilities to study coherent elastic neutrino-nucleus scattering (\CEvNS) or to detect low mass dark matter candidates. Their observation relies on the detection of nuclear recoils in the keV energy range, where energy losses result from the combination of ionization and atomic collisions. Depending on the chosen detector technology, the associated signal -- collected in the form of ionization energy, scintillation light or heat -- may differ from the one obtained from gamma calibration sources of the same energy. It is therefore of primary importance to precisely know the detector response to nuclear recoils in order to accurately reconstruct their energy.
For detectors relying only on the detection of the ionization signal, such as High-Purity Germanium Detectors (HPGe) operated at liquid nitrogen temperature, one quantity of specific interest is the dimensionless ionization quenching factor defined as the ratio of the ionization energy generated by nuclear recoils over the one generated by electron recoils of the same energy.
This quantity has been extensively measured for nuclear recoils in the 10-100\,keV range\cite{Sattler1966, Chasman1967, Shutt1992, Baudis1998, Simon2003, Benoit2007} and it follows the energy dependence predicted by Lindhard et al.\cite{Lindhard1963}.
For experiments aiming at detecting reactor antineutrinos via \CEvNS~in HPGe detectors \cite{Bonet2021d, Soma2014, Belov2015, Collar2021b}, recoils of a few \keVnr~(nuclear recoil energy) are expected, producing ionization signals in the sub-keV$_{ee}$ (electron equivalent energy) range. At these energies, precise experimental measurements of the ionization quenching factor are still lacking and the overall validity of the Lindhard theory has been questioned \cite{Sorensen2015, Sarkis2020, Sarkis2021, Barker2012}. Moreover, recent measurements have shown a deviation with respect to the prediction in silicon \cite{Chavarria2016, Izraelevitch2017} and germanium \cite{Collar2021, cdms2022}, although previous measurements \cite{Jones1971, Jones1975, Messous1995, Barbeau2007, Scholz2016} below 10\keVnr~were in agreement within uncertainties.

One of the experiments looking for the yet not observed \CEvNS~signal from reactor antineutrinos is CONUS. It consists of four 1kg-sized HPGe detectors\cite{Bonet2021d} deployed at the nuclear power plant of Brokdorf (Germany). For CONUS, the quenching factor is not only crucial for the interpretation of the Standard Model neutrino data\cite{Bonet2021}, but it is also the current main systematic uncertainty in the search for new physics \cite{Bonet2021b}.
Because of their threshold of $\sim$\,250\,eV$_{ee}$, experiments like CONUS will be able to detect the $\sim$\,1.5\keVnr~recoils induced via \CEvNS~by the highest energy reactor antineutrinos if the quenching factor is larger than 0.15. The experimental determination of this quenching factor in this range is therefore tremendously important to support the on-going and future experimental program since it determines directly the sensitivity of these experiments and their ability to perform precise measurements of the \CEvNS~signal.

This article reports a direct measurement of the ionization quenching factor of nuclear recoils in germanium for keV recoil energies. A low mass HPGe detector was positioned as an active target in monoenergetic neutron beams at the PTB Ion Accelerator Facility (PIAF) \cite{Brede1980} of the Physikalisch-Technische Bundesanstalt (PTB) in Braunschweig (Germany). Liquid scintillator (LS) based neutron detectors allowed to select nuclear recoil energies between 0.4\keVnr~and 6.3\keVnr~in the germanium target via the coincidence detection of the scattered neutrons. The article is structured as follows: after the experimental setup description in Sect. \ref{section:setup}, the analysis is detailed in Sect. \ref{section:analysis} and the quenching results are presented in Sect. \ref{section:results}. The systematic uncertainties are thoroughly discussed and as a further cross-check a comparison of the integrated recoil spectrum with a Monte-Carlo (MC) simulation is included.

All uncertainties reported in this article correspond to one standard deviation, i.e. with a coverage factor of 1.

\section{Experimental setup}\label{section:setup}

\subsection{Layout of the experiment}

\begin{figure}
    \centering
    \includegraphics[width=0.45\textwidth]{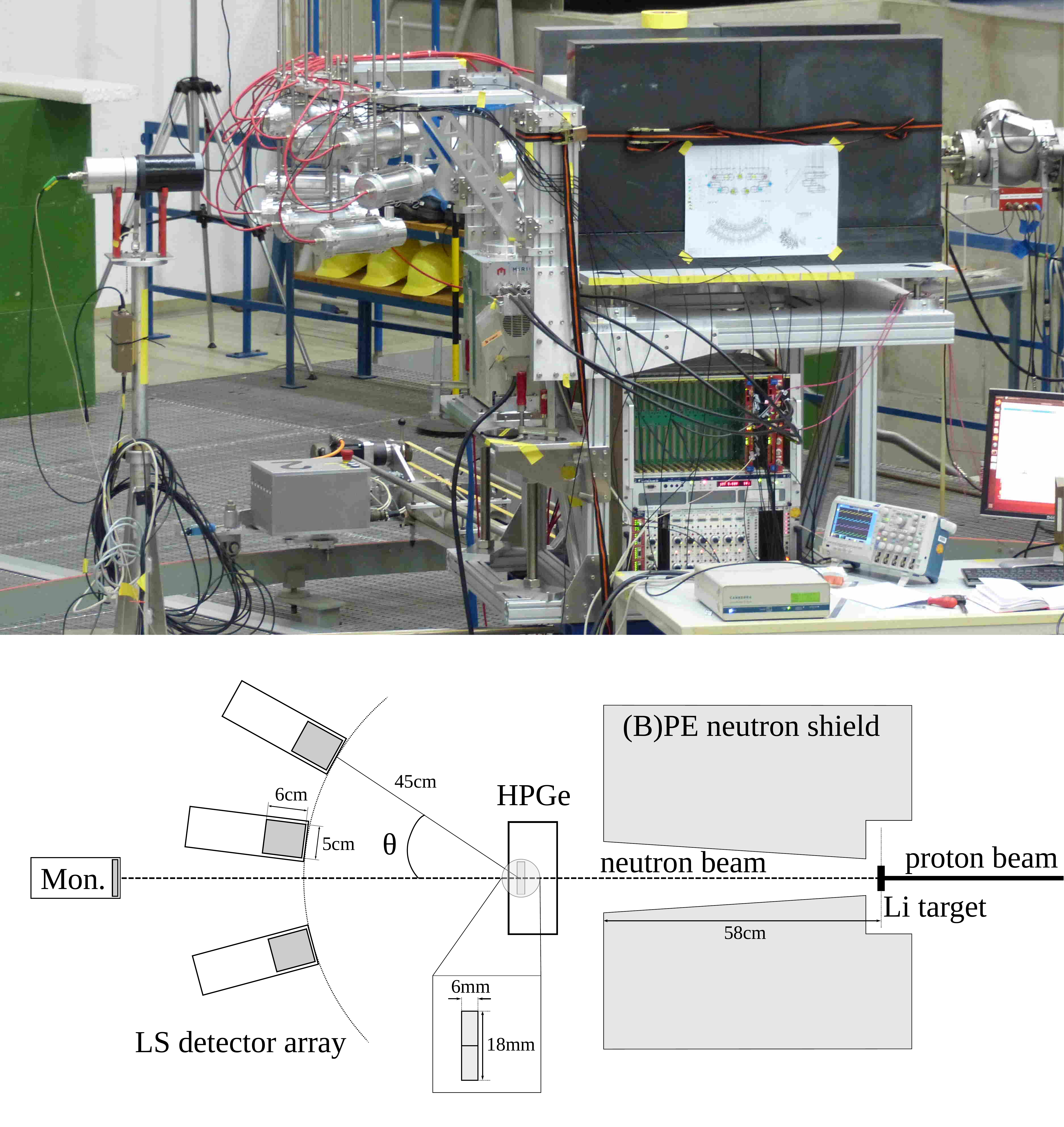}
    \caption{Photo and scheme (not at scale) of the experiment installed at the PTB facility. The proton beam is coming from the right and hits the Li target, producing a neutron beam in the forward direction. Neutrons scatter off germanium nuclei in the HPGe target detector and are detected in coincidence with the LS detectors placed behind in a half-circle. Monitoring detectors (Mon.) placed at 0 degrees allow to measure the neutron energy distribution of the beam.}
    \label{fig:setup}
\end{figure}

The experiment was set up at beam line 2 in the experimental hall of the PTB ion accelerator facility PIAF. Fig. \ref{fig:setup} shows a sketch of this setup. The HPGe detector acts as an active target deployed in the neutron beam, produced by proton-induced reactions in a lithium target. Neutrons scattered from the Ge nuclei are detected in LS detectors in coincidence with the ionization signal produced in the HPGe detector. Thus, the energy deposited in the HPGe target can be traced back using the kinematical relations for elastic neutron scattering. The LS detectors had to be shielded against the neutron source. Therefore, the lithium target was inserted into a massive shield made of borated polyethylene with a conical opening aligned with the nominal axis of the proton beam.
A lithium-6 glass scintillation detector (\textsuperscript{6}LiGl) and a barium fluoride scintillation detector (BaF\textsubscript{2}) were employed to monitor the neutron beam by time-of-flight (TOF) measurements.
The different components of this setup, their characterization and calibration are detailed in the following subsections. 

The analysis presented in this article is based on data collected in autumn 2020. Details of the runs are given in Tab. \ref{tab:runs_tof_summary}. Prior to the actual measurement of the quenching factor (runs 1 to 12 corresponding to different neutron beam energies), the LS detectors were characterized in a separate experimental campaign at another PIAF beam line (runs a, b, c).

\subsection{Neutron beams}
 
\begin{table*}
	\centering
	\begin{tabular*}{\textwidth}{ l c c c c c l r l}
	    \hline
	    run ID & $E_{p,0}$ & $m_{Li}$ & $E_{n,0}$ & $E_{n,exp}$ & $E_{n,exp}$ & $E_n$ & $\delta E_p$ & remarks \\
	    & (keV) & $\mu$g\,cm$^{-2}$ & (keV) & (keV) & (keV) & (keV) & (keV) & \\         
        & & & & $^6$LiGl & BaF$_2$ & & & \\
        \hline
        a & 2517 &  70 & 800 & $788.8 \pm 3.8$ & $793.7 \pm 7.9$ & $788.8 \pm  3.8$ & -10.7 & LS detector tests     \\
        b & 2330 &  70 & 600 & $591.3 \pm 2.8$ &                 & $591.3 \pm  2.8$ &  -8.1 &     "             \\
        c & 2238 &  70 & 500 & $483.3 \pm 2.8$ &                 & $483.3 \pm  2.8$ & -15.2 &     "             \\
        \hdashline
        1, 2 & 2240 & 100 & 500 & $488.2 \pm 2.0$ & $491.9 \pm 2.9$ & $488.2 \pm  2.0$ & -10.7 & beam test \\
        3, 4 & 2023 & 100 & 250 & $248.5 \pm 1.0$ & $248.9 \pm 1.4$ & $248.5 \pm  1.0$ &  -1.2 & quenching factor  \\
        5, 6 & 2369 & 100 & 640 &                 & $644.0 \pm 4.0$ & $644.0 \pm  4.0$ &   3.0 &     "             \\
        7, 8, 9 & 2240 & 100 & 500 &               & $490.3 \pm 3.0$ & $490.3 \pm  3.0$ &  -7.9 &     "             \\
        10, 11 & 2519 & 100 & 800 &               &                 & $790.0 \pm 11.0$ &       &     ",  adopted $E_p = 2509$\,keV     \\
        12 & 1947 & 100 & 150 &   $142.5 \pm 0.9$ & $141.1 \pm 0.9$ & $142.5 \pm  0.9$ &  -5.2 & beam test         \\
\hline
	\end{tabular*}
	\caption{Parameters of the neutron beams used for the characterisation of the LS detectors and for the measurement of the ionization quenching factors. The nominal proton and mean neutron energies are denoted by $E_{p,0}$ and $E_{n,0}$, respectively. The experimentally measured neutron energy $E_{n,exp}$ from the two employed calibration detectors is reported and the adopted value for the analysis in this article is named $E_{n}$. The difference between the proton energy $E_p$ calculated from $E_{n,exp}$ and the nominal proton energy $E_{p,0}$ is denoted by $\delta E_p$. The nominal lithium mass per unit area used for this calculation is $m_{Li}$. \label{tab:runs_tof_summary}}
\end{table*}
 
The neutron beams used for this experiment were produced by bombarding metallic lithium targets with proton beams from the 2\,MV Tandetron accelerator of PIAF. The advantage of metallic lithium target compared to the commonly used lithium fluoride targets is the increased neutron yield per unit target thickness and the reduced yield of parasitic high-energy photons from $^{19}\rm{F(p,}\alpha\gamma)^{16}O$ reactions which could cause an unwanted background increase. The proton beams were produced in pulsed mode with a repetition frequency of 1.25~MHz using the chopper/buncher system in the injection beam line. The resulting proton beam had pulse durations of about 2.5\,ns and beam currents around 1\,$\mu$A. A reference signal for TOF measurements was derived from an inductive beam pick-up located close to the neutron production target.
For proton energies between the threshold of 1.88\,MeV and 2.37\,MeV, the \textsuperscript{7}Li(p,n)\textsuperscript{7}Be reaction produces only monoenergetic neutrons. For higher proton energies, transitions to the first excited state in the residual \textsuperscript{7}Be nucleus result in a second neutron contribution of lower energy. The energy of the proton beam was measured using a 90\,\degree~analyzing magnet, which is regularly calibrated using a set of resonances and reaction thresholds.

\begin{figure}
	\centering
	\includegraphics[width=0.5\textwidth]{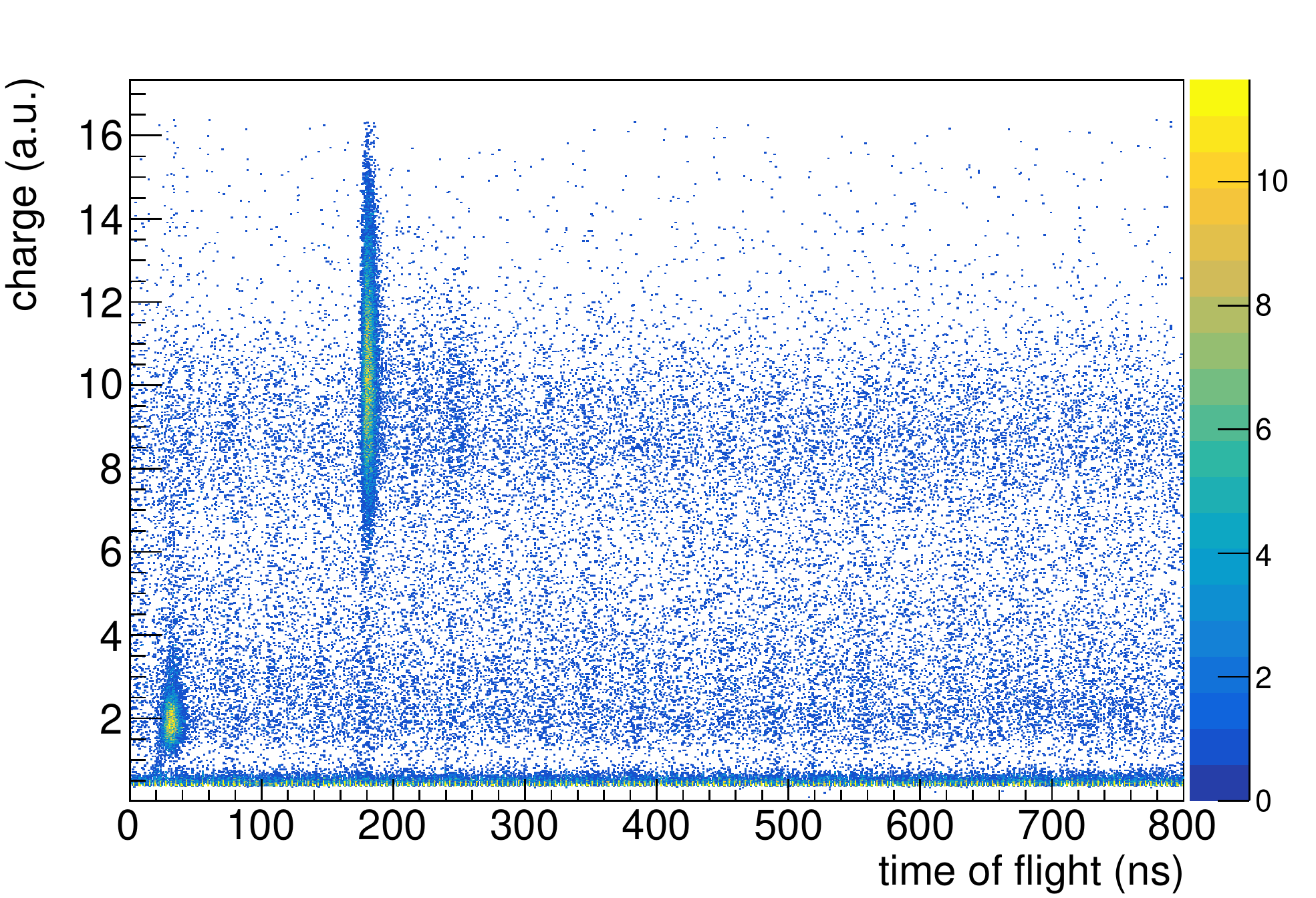}
	\caption{Pulse charge versus time-of-flight for the \textsuperscript{6}LiGl detector placed in the beam line for monitoring of the beam energy. Neutron signals can be distinguished by their high charge. In this way, they are discriminated from the ambient $\gamma$-ray background associated to a low charge.}\label{fig:tof_cuts}
\end{figure}

A direct measurement of the neutron energy distribution using the TOF method was chosen as reference for the present experiment. Two monitoring detectors were employed for this purpose: a \textsuperscript{6}LiGl detector and a fast BaF\textsubscript{2} detector. The \textsuperscript{6}LiGl detector has a diameter of 38\,mm and a thickness of only 3\,mm. The small thickness and diameter make this detector suitable for TOF measurements in the 100\,keV to 1\,MeV energy range because the crossing time is less than 0.7\,ns. In this detector, neutrons are detected via the $^6\rm{Li(n,t)}^4\rm{He}$ reaction. The large $Q$-value of 2468\,keV facilitates a separation of neutron- and photon-induced events by a simple pulse-height threshold. Fig. \ref{fig:tof_cuts} shows a two-dimensional pulse-charge versus TOF distribution for a 2240~keV proton beam corresponding to a nominal neutron energy of 500\,keV. The random background of neutron-induced events above the pulse-height threshold is mainly due to epithermal room-return neutrons. Due to the $1/v$-dependence of the cross section, the detection efficiency for these neutrons is much larger than for the fast neutrons. Without the large experimental hall at PIAF and the low-mass grid floor this background would have been much more intense.
The time resolution of the \textsuperscript{6}LiGl detector is about 4\,ns, i.e. larger than the duration of the proton beam pulses. Therefore, the fast BaF\textsubscript{2} detector (51\,mm in diameter and length, time resolution of about 0.5\,ns) was used to optimize the time structure of the beams. This detector can also detect neutrons via inelastic scattering and the detection of the deexcitation photons. Thus, it can be used to cross-check the \textsuperscript{6}LiGl measurements, but the length of the detector made it necessary to simulate the time distribution of inelastic scattering events in the detector using the MCNPX code \cite{mcnp} and include it in the analysis.
The distance between the center of the monitoring detectors and the neutron production target was measured with a laser distance meter and varied between 650\,mm and 1540\,mm with an uncertainty of 2\,mm.

\begin{figure*}
		\centering
		\begin{subfigure}[c]{0.43\textwidth}
			\includegraphics[width=\textwidth]{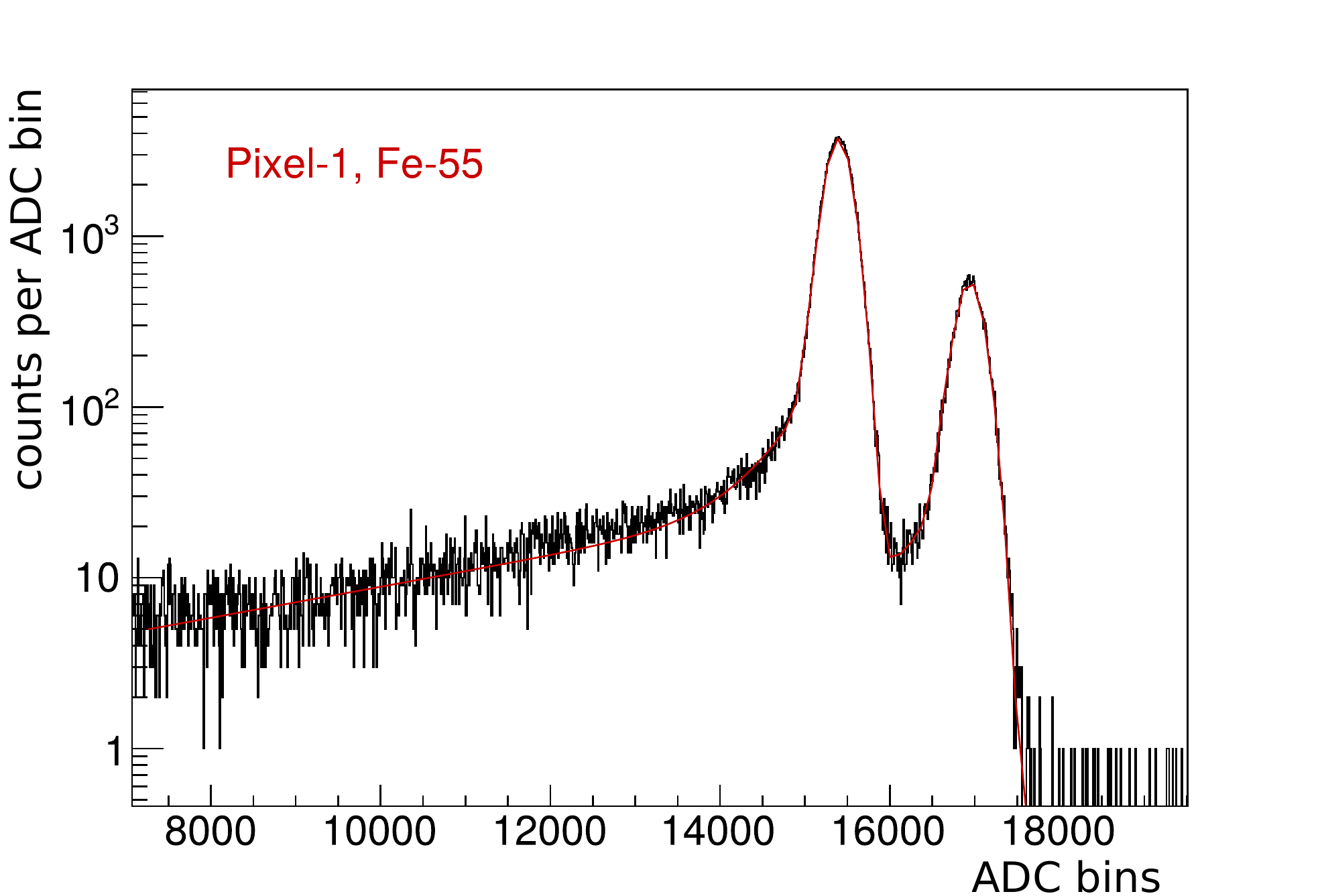}
			\caption{}\label{fig:fe55}
		\end{subfigure}
		\hspace{0.5cm}
		\begin{subfigure}[c]{0.45\textwidth}
	        \includegraphics[width=\textwidth]{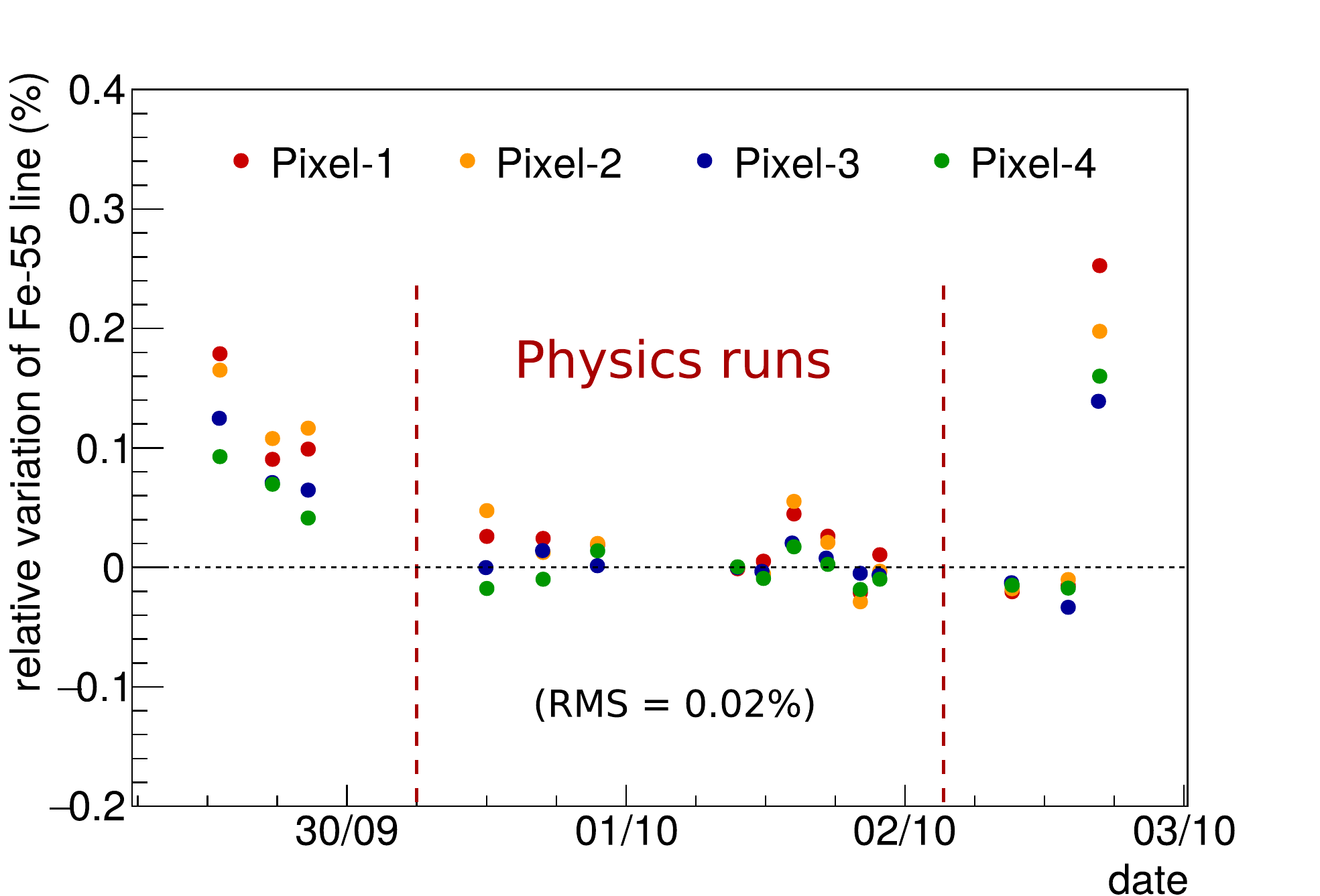}
	        \caption{}\label{fig:stability_cc_fe55}
		\end{subfigure}
		\caption{(a): Energy distribution in ADC bins for Pixel-1 from a calibration measurement with a Fe-55 source. The red line is the fit of the spectrum. The two peaks correspond to X-ray energies of 5.90 and 6.49\,keV. (b): Relative variation of the mean peak position of the Fe-55 calibration line at 5.90\,keV over the duration of the measurement campaign, relative to one of the measurements done on October 1, 2020.}
\end{figure*}

The characteristics of all neutron beams used for the present experiment are summarized in Tab. \ref{tab:runs_tof_summary}. The expected mean neutron energy $E_{n,0}$ calculated from the nominal proton energy $E_{p}$ is significantly higher than the experimental mean neutron energy $E_{n,exp}$.
$E_{n,0}$ was calculated from the proton energy by taking into account the energy loss in the target, ranging between 9\,keV and 16\,keV for the proton energies used for the present experiment. The uncertainty of $E_{n,0}$ mainly results from the uncertainty of the mass per unit area $m_{Li}$, determined via an indirect measurement using co-evaporation of lithium on a quartz balance during the preparation of the targets by physical vapour deposition (PVD) with typical uncertainties of about 10\,\%. The uncertainties of the experimental neutron energy $E_{n,exp}$ reflect the uncertainties of the peak positions of the gamma and neutron peaks in the TOF distributions, the uncertainty of the flight distance and the uncertainty resulting from the differential non-linearity of the TOF measurement. For all proton beams for which measurements with the two monitor detectors were available, the measured mean neutron energies agreed within their uncertainties. The difference $\delta E_p$ between the proton energy calculated from the experimental mean neutron energy and the nominal proton energy is about -10\,keV, except for the runs 3 to 6, where the nominal and the experimental neutron energy agreed within their uncertainties.
Similar discrepancies between the nominal and the experimental neutron energy were already observed in earlier experiments. Most likely, they are due to a not completely understood technical issue in the beam transport system at PIAF.
For the analysis of the  present experiment, the experimental mean neutron energies measured with the $^{6}$LiGl detector were adopted whenever available. For the other cases, the experimental energy measured with the BaF\textsubscript{2} detector was used. The only exception is the 800\,keV neutron beam for which no experimental measurement of the mean neutron energy was available. Here, it was assumed that the effective proton energy was 10\,keV less than the nominal proton energy of 2519\,keV, resulting in a mean neutron energy of 790\,keV for a lithium width of 100\,$\mu$g\,cm\textsuperscript{-2}. A conservative estimate of $(790\,\pm\,11)$\,keV was adopted for this dataset.

At the position of the HPGe detector 20~cm behind the exit of the collimator, the neutron beam had a diameter of about 4\,cm. The latter was estimated via a scan of the neutron counting rate along the beam axis performed with the \textsuperscript{6}LiGl detector.

\subsection{HPGe target detector}\label{subsection:hpge}
 
A dedicated n-type HPGe detector was designed in cooperation with the company Mirion Technologies (Canberra) SAS in Lingolsheim (France) to fulfill three main requirements. A small detector mass (10\,g) was chosen in order to minimize multiple neutron scattering inside the crystal which would smear out the fundamental reconstruction of the neutron scattering angle. Moreover, such a small capacitance detector garantees low noise and thus a low energy threshold. The crystal is 6\,mm thick and segmented into four quadratic pixels, each with dimensions 9$\times$9\,mm$^2$. For this geometry and neutron energies between 150\,keV and 800\,keV, the probability for multiple scattering in one pixel ranges between 10\,\% and 20\,\% only.
Then, to mitigate neutron scattering from structural materials around the detector as much as possible, the active crystal was embedded in a low-mass end-cap, arranged in such a way that the collimated neutron beam only hits the germanium crystal and two end-cap walls. These are made of very thin Be windows (300\,$\mu$m) on two opposite sides. These windows allow for low-energy calibrations with e.g. a Fe-55 source.
A typical energy spectrum of a Fe-55 source, emitting two low energy X-rays at 5.90\,keV and 6.49\,keV is shown in Fig. \ref{fig:fe55}. It demonstrates the excellent energy resolution achieved (FWHM of 135\,eV$_{ee}$ at 5.90\,keV), with a pulser resolution of about 80\,eV$_{ee}$.
The Ge crystal is cooled with an electrical cryocooling system at the nominal temperature of 88\,K.

\begin{figure}
	\centering
	\includegraphics[width=0.45\textwidth]{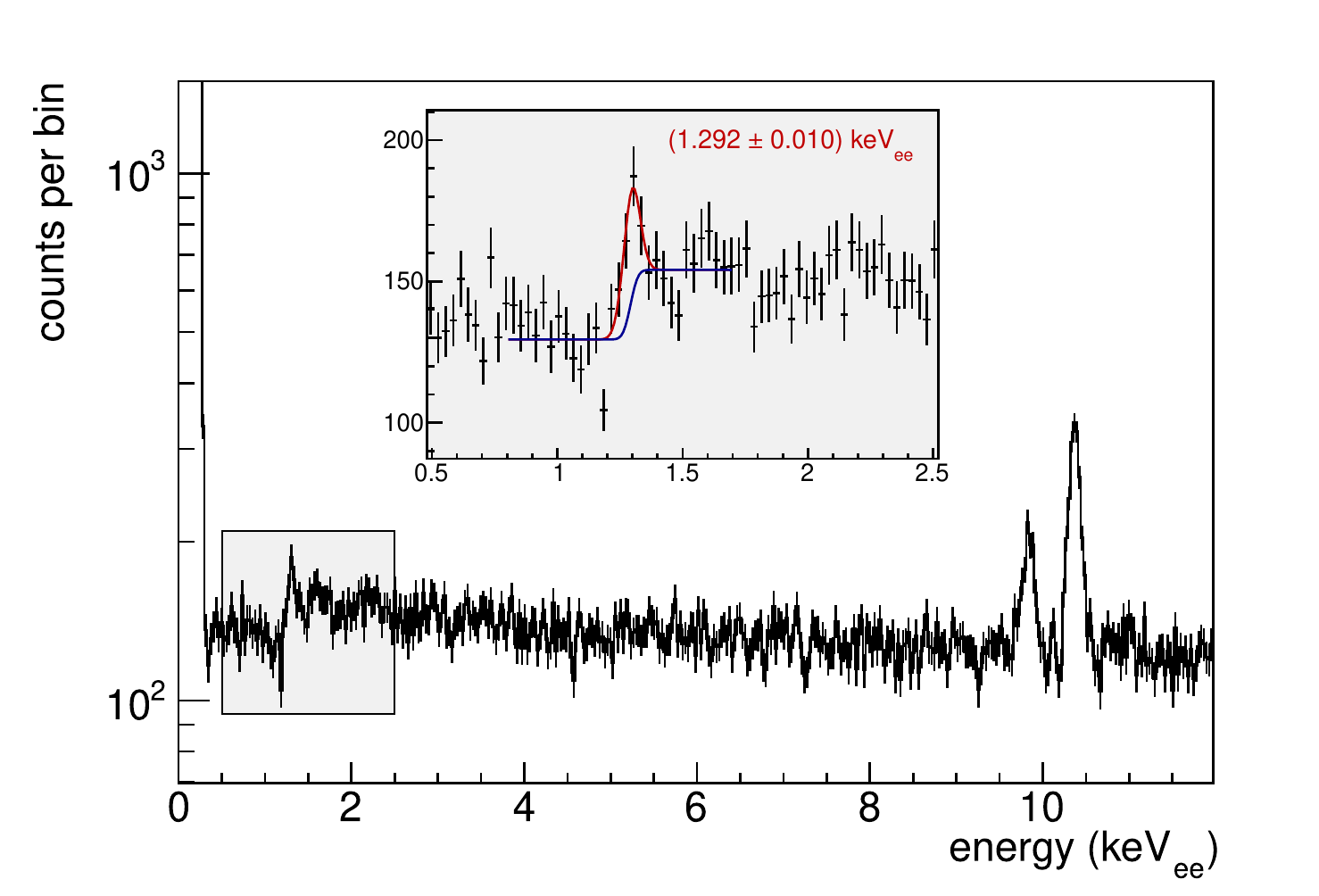}
	\caption{Calibrated background spectrum collected in the HPGe detector after beam irradiation: lines from activation are visible at 1.30, 9.87 and 10.37\keVee. The 1.30\keVee~line (red, in the inset) is used to validate the energy scale at low energy. The line position was tested with several background descriptions (blue), fit ranges and binnings.}\label{fig:background_line_1keV}
\end{figure}

Over the whole measurement campaign, the energy scale stability of the HPGe target detector was monitored via regular calibrations with a Fe-55 source. The peak position was stable below the 0.1\,\% level during the main measurement campaign as shown in Fig. \ref{fig:stability_cc_fe55}. For the physics runs used in the analysis (indicated within dashed lines), the RMS of the 9 calibration
points was used for the uncertainty on the peak positions.
After the measurement, the setup was dismantled and the detector was moved to another room to continue acquiring data to study the background produced by activation during neutron exposure. This change in experimental conditions is responsible for the small shift observed in the Fe-55 peak position in the last date cf. Fig. \ref{fig:stability_cc_fe55}. Therefore, this post-irradiation measurement was considered as an independent data set and only served as a validation of the calibration procedure.
The energy scale was determined for each pixel independently using the information of four lines: the 5.90\,keV and 6.49\,keV lines from the Fe-55, the K$_{\alpha}$ X-ray line at 9.87\,keV which can originate from fluorescent X-rays in Ge escaping from the adjacent pixels and the 10.37\,keV line from the deexcitation following electron capture (EC) in $^{68,71}$Ge, produced in the crystal by cosmic ray activation and neutron beam irradiation. The two latter intrinsic lines were observed with sufficient statistics at a rate of about a few hundred counts per day in the 10\,g crystal mass. They are visible in the background spectrum shown in Fig. \ref{fig:background_line_1keV}.
A linear calibration allowing a non-zero offset was employed, typically varying between -80 and 25\,ADC bins, and a slope of about 2600\,ADC bins/\keVee. As expected the linear fit model allows a good description of the data for all four pixels, with p-values between 0.18
and 0.95. Depending on the pixel, the statistical uncertainties of the free offsets range in (9-12)\,eV, with a mean value over all pixels of 10.6\,eV.
The L-shell line at 1.30\,keV, which originates mainly from atomic deexcitation after EC in $^{68,71}$Ge, was used only as cross-check of the calibration procedure. For this purpose, the whole background data were merged since the line was barely visible in the independent data sets and pixels because of lacking statistics.
The corresponding spectrum is shown in Fig. \ref{fig:background_line_1keV}.
With a step-like shape describing the background, the peak position is consistent with the expected value from literature within a statistical uncertainty of $\pm$\,10\,eV$_{ee}$. Changes of the binning or of the fit range do not alter this agreement, with deviations of the mean peak position staying within $\pm$\,10\,eV$_{ee}$. When using a simpler and less probable description of the background consisting in a linear model, a maximal deviation from the expected value of 12\,eV$_{ee}$ was observed.
Based on the uncertainty derived from the linear calibration fits and on this cross-check, an absolute deviation of $\pm$\,12\,eV$_{ee}$ was therefore adopted as a conservative systematic uncertainty on the energy scale.

From calibration data, the energy resolution of the detector was modeled as:
\begin{equation}\label{eq:energy_resolution}
    \sigma(E) = \sqrt{\sigma_0^2 + \mathcal{F}\cdot\varepsilon_{e-h}\cdot\text{E}}
\end{equation}
where $\sigma_0$ coincides with the pulser resolution, $\mathcal{F} = 0.13$ is the material specific Fano factor and its value was chosen such to reproduce the difference between the pulser and the Fe-55 line resolutions. For $\varepsilon_{e-h}~=~2.96$\,eV, the energy required to create an electron-hole pair at 90\,K \cite{Stuck1973}, this parametrization gives a good description of the experimental data in the region of interest up to $\sim$\,10\keVee.
The temporal stability of the energy resolution was carefully monitored throughout the experiment. HPGe detectors are indeed known to be sensitive to neutron damages. Because of incomplete charge collection, a deterioration of the resolution and the appearance of tails on the low-energy side of peaks might be observed. A critical threshold neutron fluence of about 10$^9\,$cm$^{-2}$ is often reported in the literature for n-type HPGe detectors, which are expected to be less sensitive to damages than p-type detectors\cite{Pehl1979, Raudorf1987, Descovich2005}. However, this has only been taken as a rough indication since the effect is often considered for the irradiation of large kg-sized coaxial detectors with fast neutrons (see e.g. \cite{Borrel1999}) and a dependence on the crystal temperature also seems to exist \cite{Ross2009}.
To be on the safe side, the integrated neutron fluence was therefore limited to 6$\times$10$^7$\,cm$^{-2}$ over the whole experimental campaign. The energy resolution of the 5.90\,keV reference line remained stable within $\pm\,3$\,eV$_{ee}$ during the entire measurement campaign. Moreover, the 59.5\,keV line from regular measurements with an Am-241 source was used to monitor the energy resolution at higher energy, where effects due to incomplete charge collection become more visible due to the dominance of the second term in Eq. (\ref{eq:energy_resolution}).
The width of the line remained stable with a resolution (FWHM) of 416\,eV$_{ee}$ at 59.5\,keV and did not show any indication of a degradation of the charge collection during and after the neutron irradiation.

\begin{figure}
	\centering
	\includegraphics[width=0.4\textwidth]{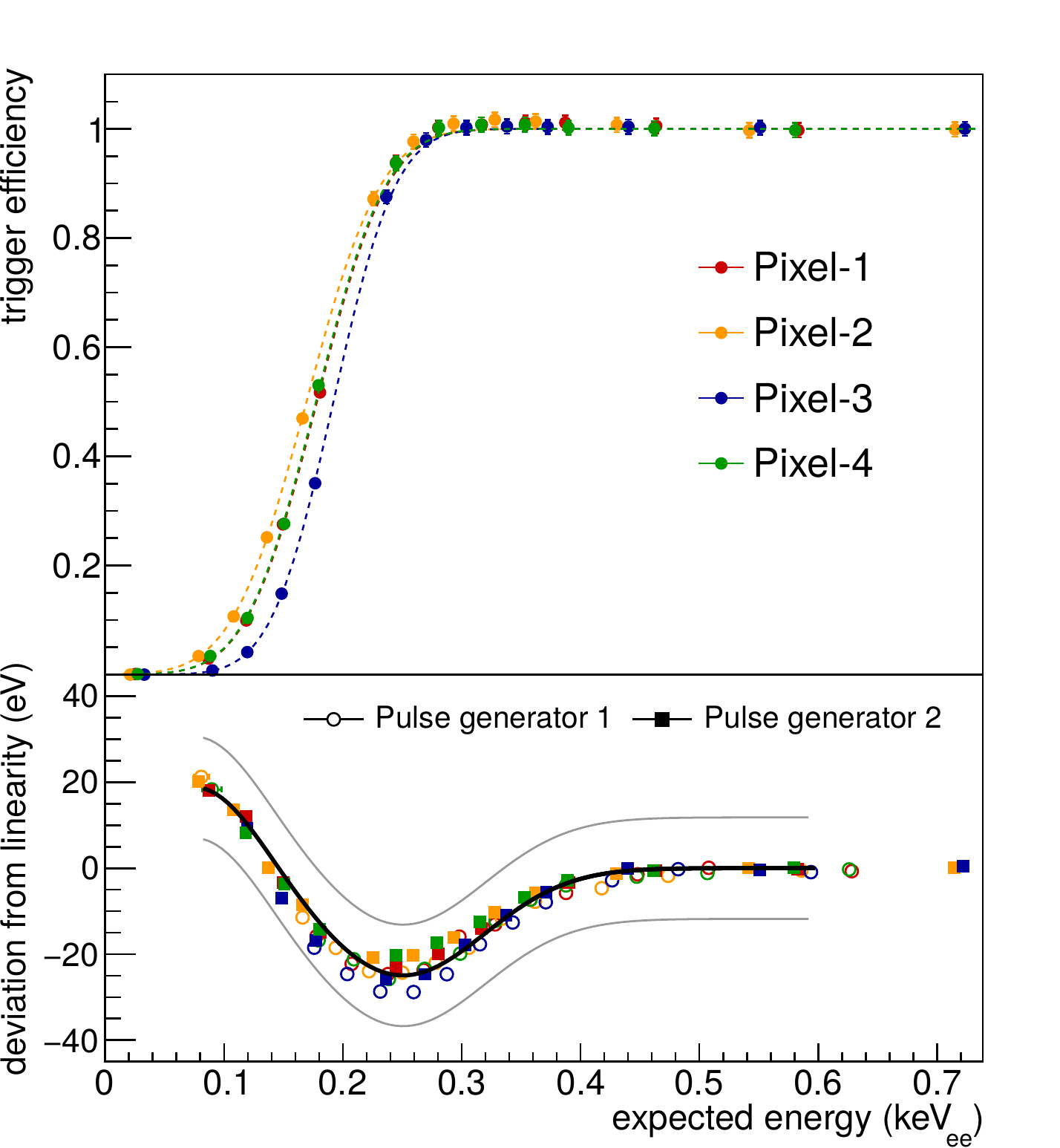}
	\caption{Top: measured trigger efficiencies for the four pixels (points) and best-fit model (dashed lines) using the analytical description of Eq. (\ref{eq:trigger_efficiency}). Bottom: deviations from a purely linear energy scale measured with two different pulse generators (points) described by the black line. The $\pm$\,12\,eV$_{ee}$ uncertainty band derived for the energy scale is shown in gray and it largely covers the small variations observed between the measurements and pixels. Both effects are included in the detector response matrix used to compute the expected energy distributions (cf. Sect.\,\ref{subsection:analysis})}\label{fig:trigger_efficiency}
\end{figure}

Since no photon sources are available for calibration purpose between the threshold and 1.30\keVee, detailed investigations of the detector response were performed with artificial signals produced by pulse generators with the same risetime as for physical signals. Fine-grained scans allowed to precisely measure the trigger efficiency (cf. Sect.~\ref{subsection:daq}) for the four pixels as a function of the expected energy, derived from a careful calibration of the pulse generator amplitudes. As shown on the upper figure in Fig. \ref{fig:trigger_efficiency}, they exhibit a similar behaviour. A very good description of the experimental trigger efficiency curves was obtained via:

\begin{equation}\label{eq:trigger_efficiency}
    \varepsilon_{trig} = 0.5\cdot\left(1+\text{erf}\left(\frac{E_{ee}-t_1}{t_2}\right)\right)
\end{equation}
where erf is the error function with typical parameters $t_1$~=~170\,eV$_{ee}$ and $t_2$~=~65\,eV$_{ee}$.

With the same pulser scans, the linearity of the DAQ chain in the sub-\keVee~region was investigated. Deviations might appear at low energy in dependence of the reconstruction algorithms \cite{Soma2016}. Such deviations from a purely linear behavior were observed below $\sim$\,400\,eV$_{ee}$ and are shown in the bottom panel of Fig.~\ref{fig:trigger_efficiency}. They were attributed to two nearly independent DAQ-related effects. First, as the amplitude of the signals decreases, artificial delays of the trigger time stamp up to a few $\mu$s were observed. This effect is due to the use of a specific trigger logic (cf. Sect. \ref{subsection:daq}) and is observed in quenching data as well, see Fig. \ref{fig:correlation_liq_ge}. Because of the relatively short shaping time of 4\,$\mu$s, these delays affect the reconstructed energy value obtained by trapezoidal shaping. The corresponding negative non-linearity effect was less than 5\,eV$_{ee}$ at 400\,eV$_{ee}$ and reaches its maximum of about 25\,eV$_{ee}$ at 250\,eV$_{ee}$. Second, the drop of the trigger efficiency for $\lesssim\,$250\,eV$_{ee}$ is responsible for an artificial positive deviation from linearity, overwhelming the above-mentioned effect. This will be further discussed in the analysis section (Sect.\,\ref{subsection:analysis}) and is illustrated in Fig. \ref{fig:trigger_effect}.

Being precisely measured, all above mentioned effects are corrected for in the quenching analysis (Sect.\,\ref{subsection:analysis}) by taking them into account in the form of a detector response matrix. The energy resolution is implemented following Eq.~(\ref{eq:energy_resolution}). For the trigger efficiency, the analytical description of Eq.~(\ref{eq:trigger_efficiency}) is used and the non-linearies are described by the black model shown in Fig.~\ref{fig:trigger_efficiency}. This matrix is then used to compute the expected energy distributions.


\subsection{Neutron detectors}\label{subsection:ls}
For the detection of the scattered neutrons, eleven LS detectors were designed and assembled at MPIK. They consist of cylindrical PTFE cells with very thin front walls. Diameter and height of the cells are 5 and 6\,cm, respectively, resulting in an active volume of about 120\,cm\textsuperscript{3}. The cells were filled with EJ-301(Eljen Technology) LS, which is identical to the well-known NE213 scintillator \cite{Batchelor1961} and was chosen for its fast response, high light yield and good pulse shape discrimination (PSD) properties.
The cells are optically coupled to 2" photo-multiplier tubes (PMTs). Detector cells and PMTs are embedded inside a light-tight aluminium housing. Neutrons are detected via the scintillation light produced from the prompt recoil signal, providing a fast response ($\sim$\,ns) adapted for the timing requirements of the coincidence measurement.

\begin{figure}
	\centering
	\includegraphics[width=0.45\textwidth]{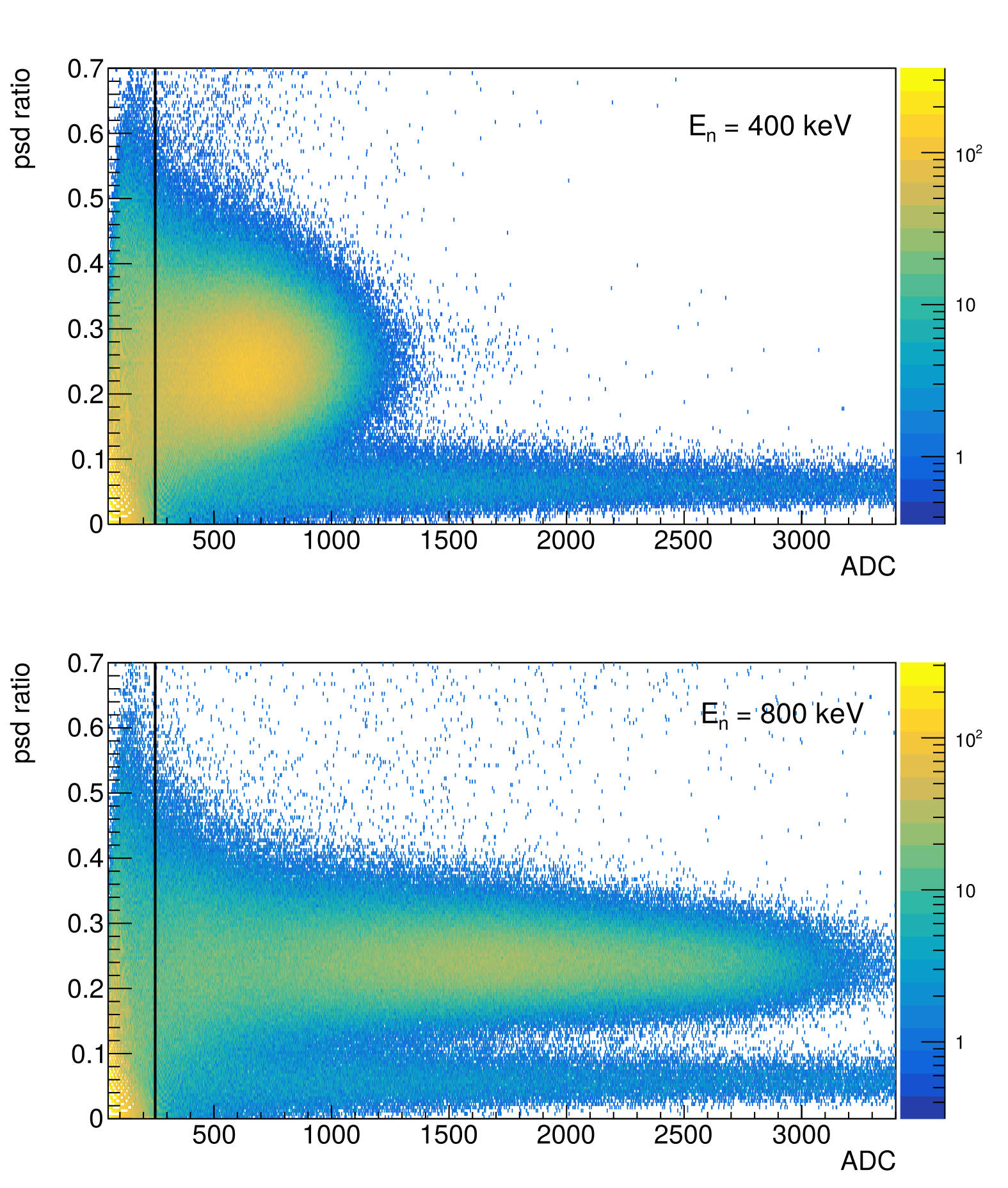}
	\caption{PSD ratio versus charge (in ADC units) for a LS detector placed in the neutron beam line during commissioning. The black line indicates the threshold used for the analysis. The neutrons, inducing proton recoils in the liquid, are identifiable by their high PSD ratio ($\gtrsim$\,0.1). The lower band consists of interactions from ambient $\gamma$-rays.}\label{fig:psd_liq}
\end{figure}

The neutron response of these detectors was characterized in detail in a dedicated commissioning campaign to ensure that all the detectors were equivalent and their data can be combined. The measurements were carried out in several monoenergetic neutron fields produced in open geometry using the \textsuperscript{7}Li(p,n)\textsuperscript{7}Be reaction and a 70\,$\mu$g/cm\textsuperscript{2} metallic lithium target and ns-pulsed proton beams (cf. Tab. \ref{tab:runs_tof_summary}). For neutron emission angles of 0°, the neutron fluence at the position of the detector cells was measured using a de Pangher \cite{Pangher1966} long counter which is the reference instrument for routine fluence measurements at the PTB. For non-zero emission angles, the neutron fluence was calculated from the measured 0° value using the known angular dependence of the neutron yield \cite{Liskien1975, NeuSDesc}. Non-monoenergetic neutrons of lower energy resulting from neutron scattering in the production target were discriminated using the time-of-flight (TOF) technique. Residual background resulting from forward scattering neutrons from air and structural materials was determined in a separate measurement with a shadow cone placed between the production target and the detectors. The trigger threshold was calibrated in electron-equivalent energy using Cs-137 and Ba-133 photon sources emitting $\gamma$-rays in the range from 80\,keV to 667\,keV. For a trigger threshold of 12\,\keVee, the neutron detection efficiency of the detectors was about 75\,\% over the energy range of interest from 250\,keV to 800\,keV with a slight variation of about 5\,\%.
Neutron- and photon-induced events were separated by measuring the signal decay times using the ratio of the signals integrated over a short and a long gate period. Representative PSD distributions for two different neutron beam energies are shown in Fig. \ref{fig:psd_liq}. A good discrimination between proton recoils (induced by neutrons) and electron recoils (ambient $\gamma$-rays) is obtained and allows to get rid of about 80\,\% of the ambient $\gamma$-ray background while keeping a neutron efficiency of 85-95\,\% depending on the beam energy.

\begin{figure}
	\centering
	\includegraphics[width=0.45\textwidth]{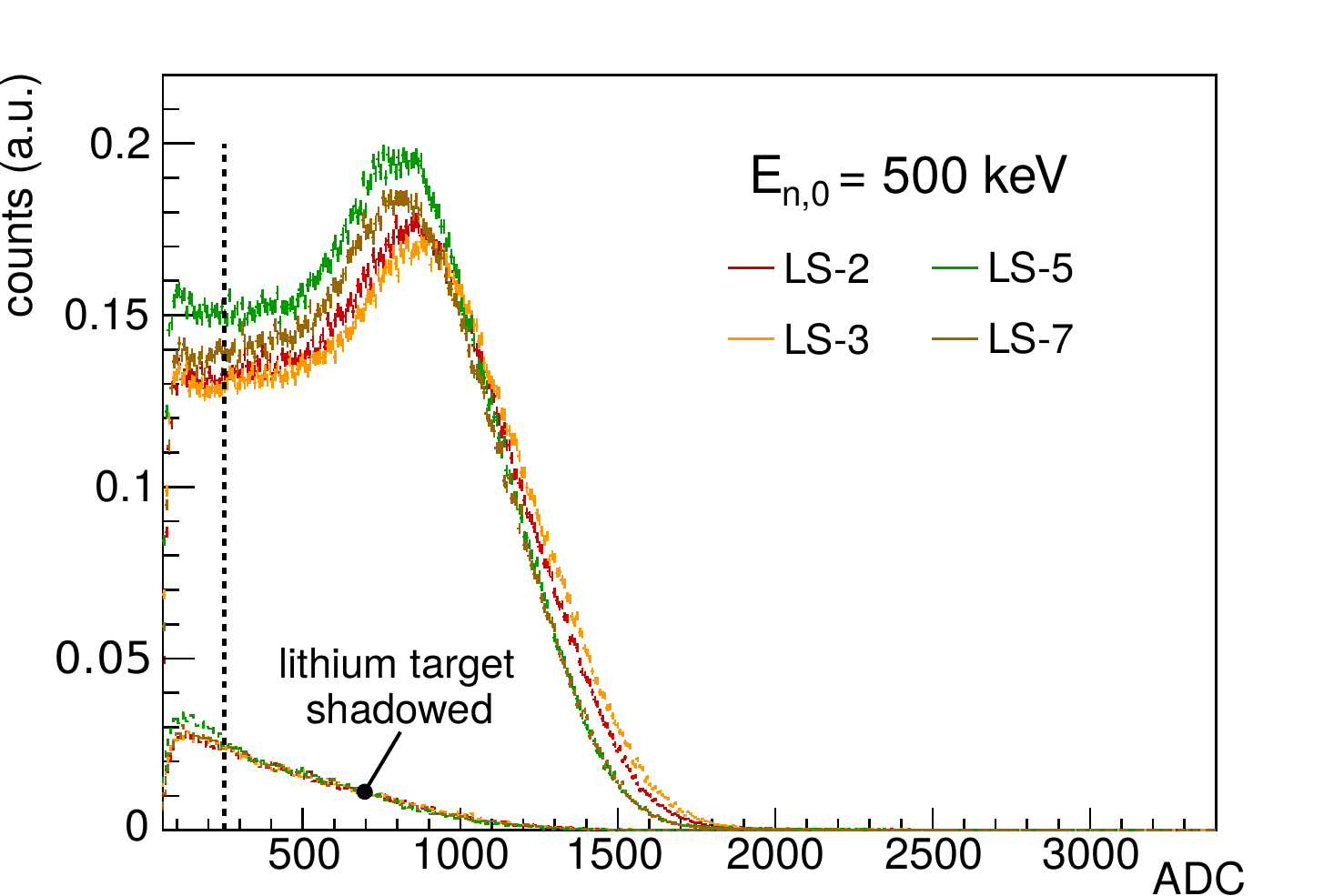}
	\caption{Energy distribution obtained for four representative LS detectors exposed to a 500\,keV neutron beam during commissioning. The bottom distributions are obtained by shadowing the LiF target with a polyethylene cone and allow to estimate the background contribution from air scattering. The threshold at 250\,ADC used for the analysis (black dashed line) was chosen such to minimize the discrepancies in terms of neutron efficiency between the detectors.}\label{fig:liq_spectra}
\end{figure}

Based on these commissioning data, the small discrepancies observed between the eleven detectors and variations over time were quantified in terms of relative variations in the neutron detection efficiency. The impact of small shape discrepancies in the neutron response distribution between the detectors -- shown in Fig. \ref{fig:liq_spectra} --  was mitigated to 3\,\% by choosing an energy threshold of 250\,ADC bins corresponding to 12\,\keVee, indicated by a dashed line. During the measurement campaign, the stability of the LS detectors, including scintillator liquid, high-voltage, PMT and trigger thresholds, was regularly monitored using the Cs-137 and Ba-133 photon sources. The small gain variations between the detectors amount to less than 1.5\,\%. Finally, time variations in the response are responsible for an additional 1.5\,\% effect. 

\begin{figure}
	\centering
	\includegraphics[width=0.45\textwidth]{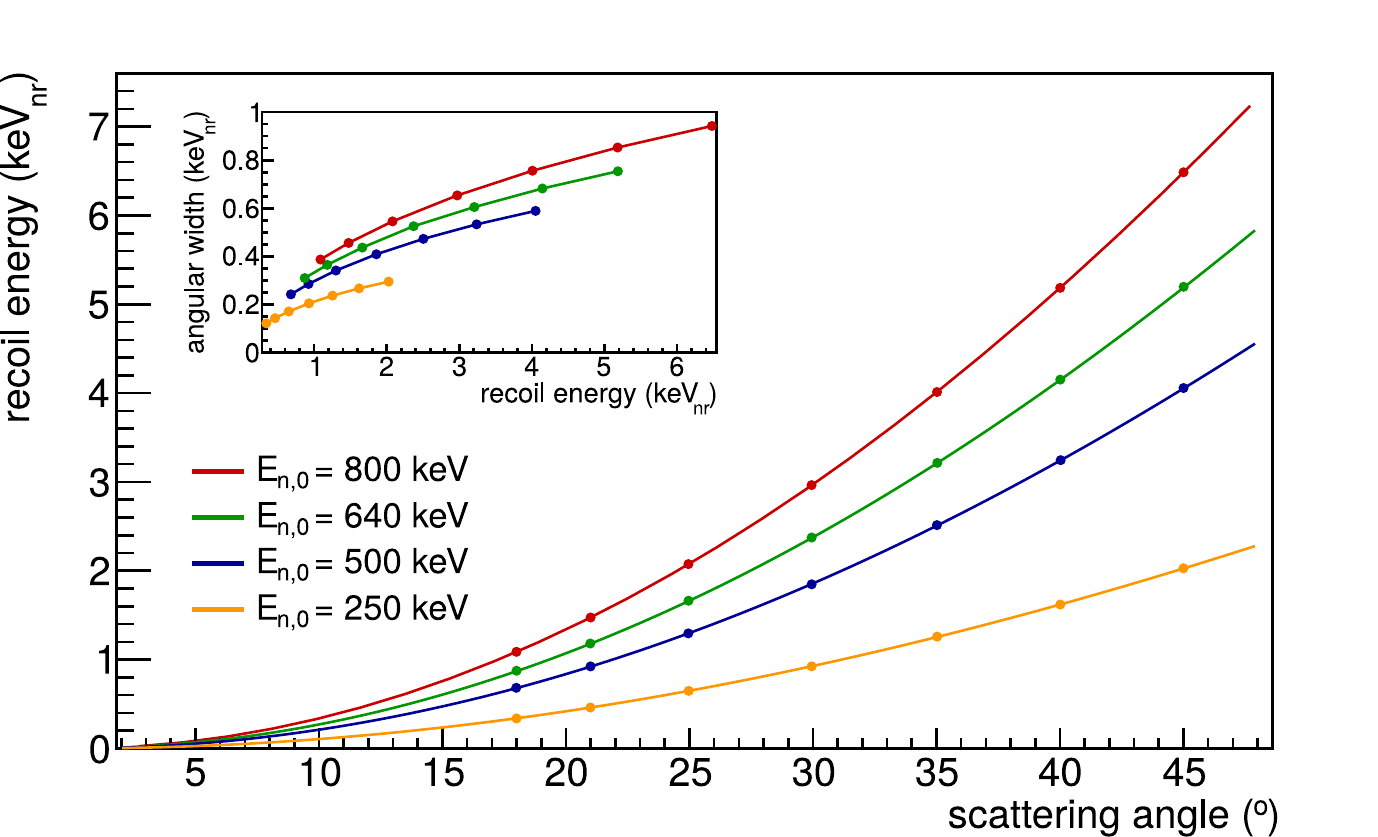}
	\caption{Nuclear recoil energy obtained in germanium for incoming neutrons with energies 250 to 800\,keV as a function of their scattering angle, according to Eq. (\ref{eq:enr_f_theta}). The expected angular spreads for the chosen setup are represented in the inset. The dots indicate the angles chosen for the measurement.}\label{fig:beam_energy}
\end{figure}

The detectors were positioned in a circular array at about 45\,cm away from the HPGe target with a better coverage for the smallest angles -- corresponding to the lowest recoil energies. The nuclear recoil energy $E_{nr}$ deposited in the germanium target of atomic mass A for a neutron scattering angle $\theta$ is:
\begin{equation}\label{eq:enr_f_theta}
E_{nr} = \frac{2E_0}{(A+1)^2}\left( A+\sin^2(\theta)-\cos(\theta)\cdot\sqrt{A^2-\sin^2(\theta)}\right)
\end{equation}
where $E_0$ is the initial neutron energy and $\theta$ is the scattering angle of the neutron as indicated in Fig. \ref{fig:setup}. Scattering angles between 17\degree~and 45\degree~were selected, which allowed to probe nuclear recoils between 0.4 and 6.3\keVnr, as illustrated in Fig. \ref{fig:beam_energy}.

\subsection{DAQ}\label{subsection:daq}

A DAQ system based on commercially available CAEN electronic modules was used to acquire data from the LS and the HPGe detectors as well as the beam information. In order to fulfill both the requirements of a TOF analysis with fast PMT signals and of the slower shaping times used for the HPGe detector signals, a modular configuration was developed from digitizers offering pulse processing implemented at the FPGA level.

The HPGe signals were sampled at a rate of 100\,MHz rate by a V1782 multi-channel analyzer. A combination of a slow and a fast triangular discriminator was used for the trigger. This gave better results in terms of detection efficiency of small signals compared to the standard algorithm. The fast and slow triangular discriminators had shaping times of 0.8\,$\mu$s and 2.4\,$\mu$s, respectively. The energy of each event was reconstructed by a trapezoidal shaping filter with a shaping time of 4\,$\mu$s, optimized in terms of energy resolution.

The signals from the PMTs of the LS detectors were acquired with a V1725 module at a sampling rate of 250\,MHz. The charge integration windows were chosen to maximize the particle discrimination by PSD. A time resolution of 6\,ns (FWHM) for physical LS signals was achieved.

The two modules shared a common synchronization clock, allowing to identify coincidences between LS and HPGe events. For practical reasons, the reference signal provided by the proton beam pick-up was sampled by both modules and was saved only in the presence of a physical trigger (LS or HPGe).

All triggering events were saved and the coincidence selection between the signals from the HPGe and the LS detectors was performed offline. Typical rates encountered during a quenching data run were in the range of (250-550)\,Hz for the LS detectors. For the HPGe target, typical rates were always dominated by the noise trigger rate, ranging from 60\,Hz to 1.2\,kHz depending on the pixels.

\section{Analysis}\label{section:analysis}

\subsection{Coincidence data selection}

\begin{figure}
	\centering
	\includegraphics[width=0.45\textwidth]{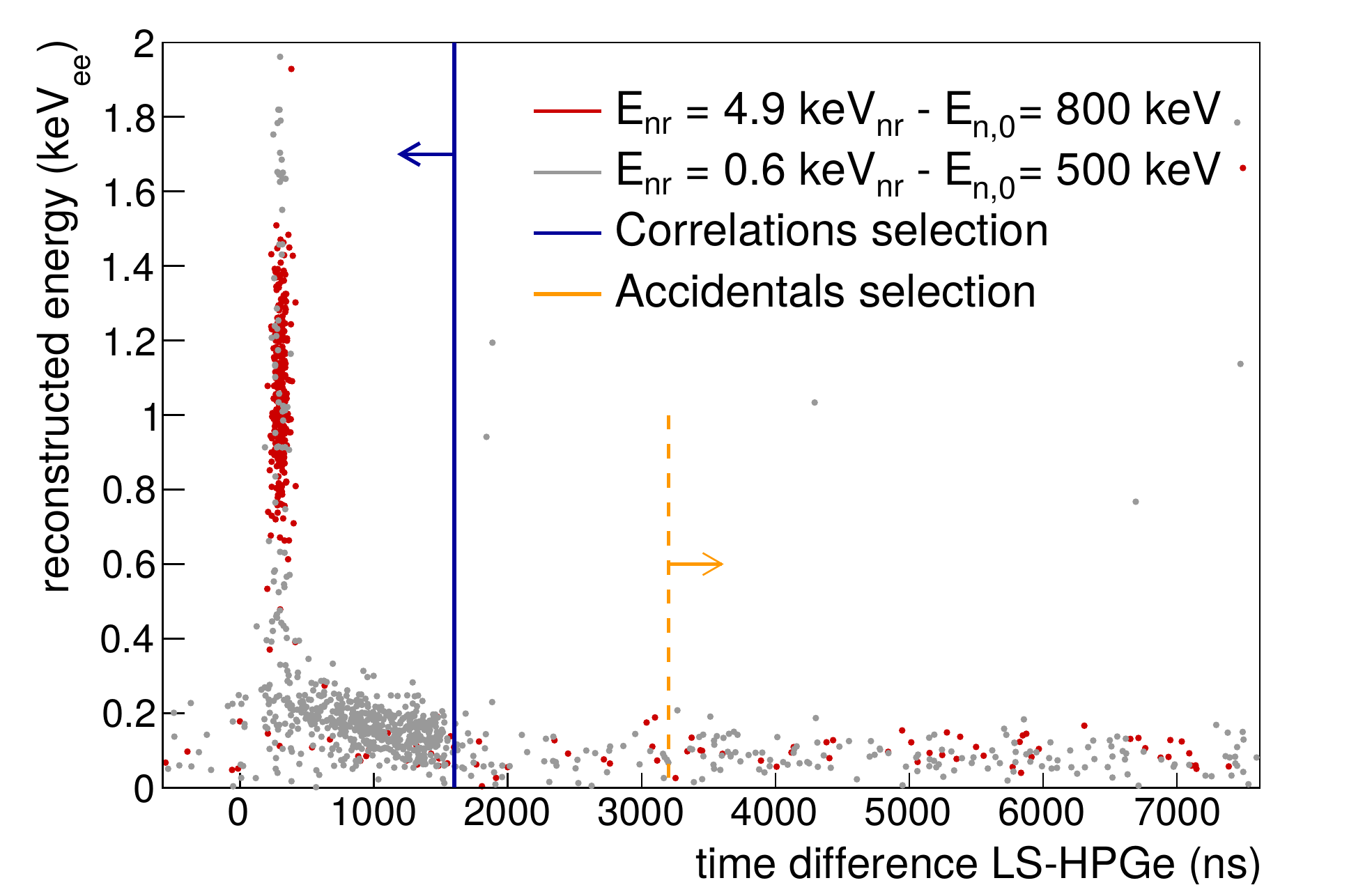}
	\caption{Data selection for the coincidence analysis based on the time difference between a LS and a HPGe signal for two different expected nuclear recoil distributions (red at 4.9\keVnr~and gray at 0.6\keVnr). The energy reconstructed in the HPGe detector is represented on the y-axis. Correlated signals are selected for $\Delta$T$<1.6\,\mu$s (blue line) enclosing the smearing of the distributions at low energy, whereas accidental coincidences are estimated in 10 windows starting from $\Delta$T$>3.2\,\mu$s (yellow line).}\label{fig:correlation_liq_ge}
\end{figure}

The data selection for the analysis relies on a three-fold time coincidence between the proton beam pick-up signal, the signals from the HPGe target and the LS detectors. In practice, this was achieved by first selecting the neutron-induced events in the LS detectors using their TOF and PSD distributions. A selection window of 20\,ns was chosen for the TOF distribution. The size of this window was determined from the LS data without coincidence requirements and corresponds roughly to a 3\,$\sigma$ wide time window around the neutron arrival time. This selection was extended to 40\,ns for the 250\,keV data set. Neutron signals were discriminated from the ambient $\gamma$-ray background with an additional PSD selection. In order to keep a constant neutron detection efficiency, the PSD cut was derived individually for each LS detector from a reference calibration measurement with an AmBe neutron source. Events with a PSD higher than q$_{cut}$ = q$_{n} - 3\sigma_n$ were selected, where q$_{n}$ and $\sigma_n$ are the mean peak position and standard deviations of the nuclear recoil PSD distributions. The time window for a coincidence between a HPGe event and a LS event was fixed to 1.6\,$\mu$s to take into account the increased smearing of the trigger time stamp at low energies detailed in Sect. \ref{subsection:hpge}. Accidental background essentially consists of random noise. Its contribution was evaluated by opening 10 random coincidence gates outside of the main coincidence window. This selection is illustrated in Fig. \ref{fig:correlation_liq_ge}. In this way, a coincidence rate of the order of 100 counts per hour and per HPGe pixel was obtained for each angle. The coincidence energy distributions of interest for the quenching analysis were obtained by adding up the contributions of the four pixels.

\subsection{Analysis procedure}\label{subsection:analysis}

The expected nuclear recoil energy distribution for each angle is obtained by running a MC simulation taking into account the geometrical extension of the detectors, the mean free paths of the neutrons inside the LS and the width of the neutron beam. The mean nuclear recoil $E_{nr}$ for each detector position is taken as the mean of the distributions. The widths of the expected distributions were found to be systematically smaller than the ones observed in the experimental quenching data. This effect is further discussed at the end of this section. Due to the lack of a quantitative description of this spread, a conservative approach was adopted and the nuclear recoil energy distributions were simply modeled by a Gaussian with mean $E_{nr}$ derived from the MC distribution and a free width $\sigma_{nr}$ (in \keVnr).
An uncertainty on $E_{nr}$ is estimated for each angle from the propagation of the experimental uncertainties of the spatial coordinates measurement on site. An angular uncertainty of 0.5\,$^{\text{o}}$ to 1.5\,$^{\text{o}}$ was obtained, which translates into a 0.05\keVnr~to 0.2\keVnr~uncertainty in nuclear recoil energy, depending on the angle and on the beam energy.

\begin{figure}
	\centering
	\includegraphics[width=0.45\textwidth]{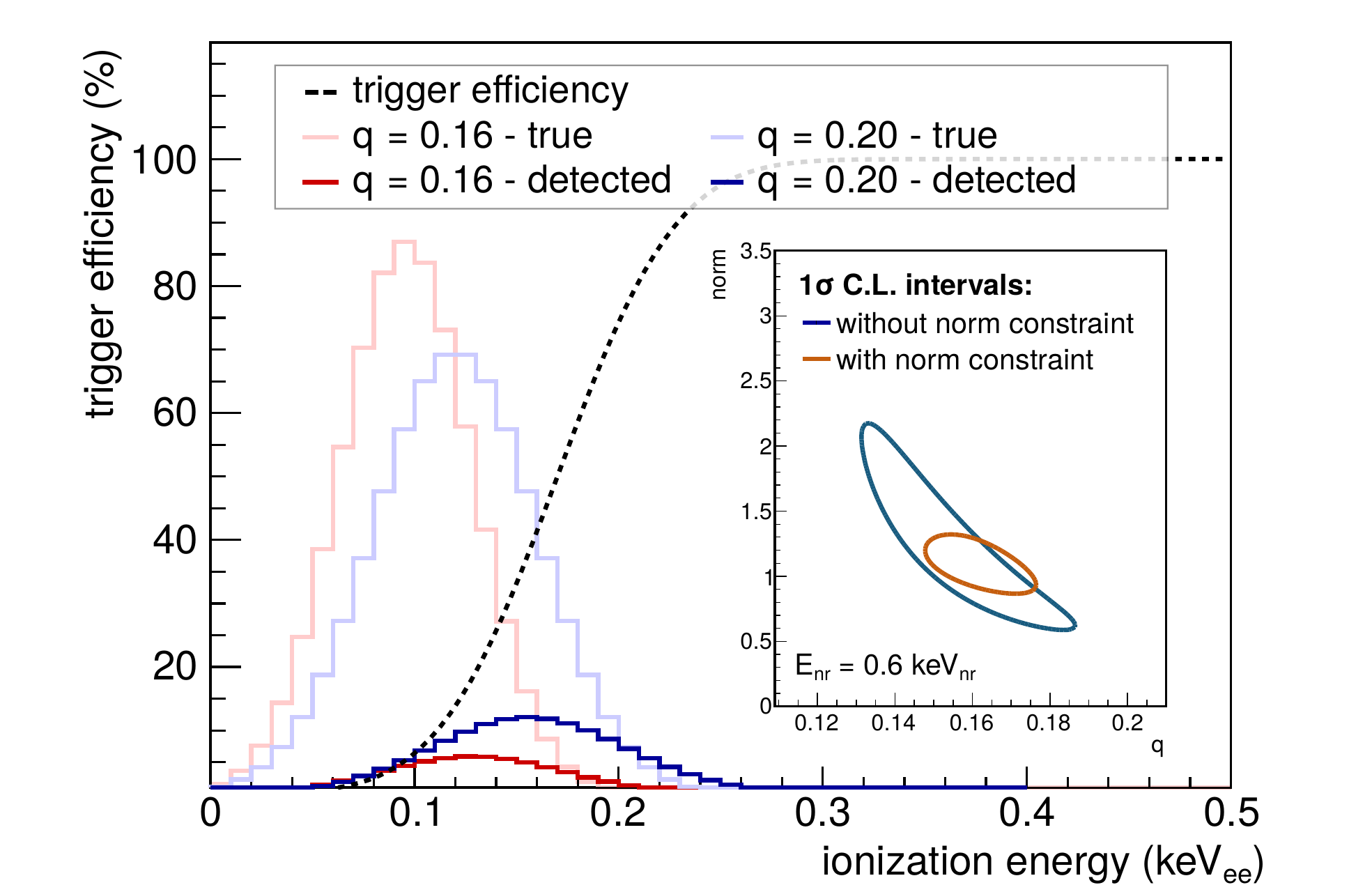}
	\caption{Illustration of trigger efficiency effects for an initial recoil energy of 0.8\,keV$_{nr}$ under the assumption of an ionization quenching factor q of 0.16 (red) or 0.20 (blue). The "detected" distributions are obtained by convolving the "true" distributions with the HPGe detector response, including the energy resolution and the effects of trigger efficiency and energy scale non-linearities discussed in Sect.~\ref{subsection:hpge}. In particular, if the trigger efficiency is not taken into account properly, this can lead to apparent non-linearities in the energy scale and biased results in terms of ionization quenching factor. Due to the poor discrimination between the detected shapes, rate and ionization quenching factor strongly anti-correlate at these energies. This is illustrated by the blue 1\,$\sigma$ contour in the inset plot obtained when fitting the low energy coincidence data distribution without any norm constraint. The addition of the norm constraints summarized in Tab. \ref{table:norm_constraint} clearly provides a sensitivity improvement and allows to reduce the statistical uncertainties on q by about a factor 2 as shown by the corresponding orange contours.\label{fig:trigger_effect}}
\end{figure}

These nuclear recoil energy distributions are convoluted with the response matrix of the HPGe detector to obtain the model $\mathcal{M}_{ee}$ -- including the quenching process -- used to describe the data.
On top of resolution effects, it is of primary importance to take into account the detector response at low energies detailed in Sect. \ref{subsection:hpge}. In particular, detected distributions in the region below 1\keVnr~are expected to be artificially shifted towards higher energies due to the drop of trigger efficiency. If the effect is not properly considered, this may lead to a biased value of the ionization quenching factor as illustrated in Fig. \ref{fig:trigger_effect}.

In order to normalize the models $\mathcal{M}_{ee}$ to the data, a common signal coincidence rate $n^d_0$ per unit of beam charge was determined for each data set. Indeed, within a given data set $d$ corresponding to a given neutron beam energy, the coincidence rate obtained for each angle is expected to reflect only the differential angular elastic neutron scattering cross-section in Ge. It was cross-checked that the correlated counts corrected by the neutron elastic scattering cross-section in Ge and the acquisition time for each angle were constant within a data set for the energies above the region of trigger efficiency loss. The common rate $n^d_0$ was taken as the mean of the obtained values.

For the modeling of the background, the accidental component is directly estimated from data as described in the previous paragraph. An additional continuous correlated component arising from multiple scattering in the crystal and the surrounding materials populates the correlated energy distributions up to a few keV. It amounts to about 20\,\% of the total correlated rate between 0 and 12-13\,\keVee~(end of the dynamic range). For the 500, 640 and 800\,keV datasets, this background contributes to $(1-4)$\,\% to the counting rate in the region of interest defined as a 2\,$\sigma$ window around the signal peak. It is higher for the 250\,keV dataset with a contribution at the level of $\sim$\,10\,\%. It has therefore been modeled for each dataset $d$ by a simple flat contribution of common amplitude $b_0^d$ obtained by a combined fit of the coincidence distributions above the peak region. Two refined modelisations of this background were also tested as extreme cases: the first one with a quadratic increase extracted from the integrated recoil distributions (see Fig.~\ref{fig:integrated_spectra} in Sect.~\ref{subsection:integrated_recoils}) and the second one with the same model convoluted by the trigger efficiency. The impact on the result was found to be negligible.

The coincidence energy distributions for each angle $i$ were treated separately with a model-independent ionization quenching factor $q_i$ -- assumed to be constant in energy -- by minimizing the following $\chi^2$ function:

\begin{table}
	\centering
	\begin{tabular}{l | c }
		source of uncertainty & \\
		\hline
		cross-section & 5.0\,\% \\
		LS stability & 1.5\,\% \\
		deviation between LS detectors & 1.5\,\% \\
		shape variations between LS detectors & 3.0\,\% \\
		\hline
		total & 6.2\,\%
	\end{tabular}
	\caption{Constraints in terms of coincidence rate used in the fit. The constraint on the scattering cross-section reflects the discrepancies between the neutron databases. Constraints concerning the LS detectors are derived from the characterization detailed in Sect.  \ref{subsection:ls}.}\label{table:norm_constraint}
\end{table}

\begin{align}
\chi^2_i =& \sum_{j=0}^{N_{bins}} \frac{[d_j - n^d_i\,c_i\,f_{\theta_i}\,\mathcal{M}_{ee,j}(q_i, \sigma_{nr,i})- c_i\,b_{0}^d - a_j^i]^2}{\sigma_j^2} \nonumber \\
&+ \frac{[n^d_i - n^d_0]^2}{\Phi_d^2}\label{eq:chi2_pull}
\end{align}
where the index $j$ is running over the bins of the coincidence energy distribution $i$ and $d_j$ are the corresponding counting rates with their associated statistical uncertainties $\sigma_j$. 
Apart from the ionization quenching factor $q_i$ and the width $\sigma_{nr,\,i}$ (expressed in units of \keVnr), which are left free in the fit, the expected model $\mathcal{M}_{ee,\,j}$ depends on all detector model ingredients described in Sect. \ref{subsection:hpge} such as trigger efficiency and energy non-linearities. It is scaled using the integrated beam current $c_i$ and $f_{\theta_i}$, a dimensionless factor accounting for the differential angular elastic neutron scattering cross-section. Also included are the contributions from the accidental coincidences ($a_j^i$) and the correlated background ($b_0^d$).
The nuisance parameter $n^d_i$, driving the normalization of the model, is constrained in the fit via the addition of a Gaussian pull-term incorporating the normalization knowledge: $n^d_i$ can vary around its nominal value $n^d_0$, with a tolerance driven by $\Phi_d$, coming from both the characterized differences in terms of neutron detection efficiency of the LS and the knowledge on the cross-section. For the latter, the following evaluated nuclear data libraries were considered: ENDF/B-VII\cite{endfb7}, JEFF-3.3\cite{jeff33}, JENDL-4.0\cite{jendl} and CENDL-3.1\cite{cendl}. The corresponding uncertainty on the expected counting rate was taken from the dispersion between the different evaluated databases.
These normalization constraints are summarized in Tab.~\ref{table:norm_constraint}. They introduce sensitivity to the particularly delicate low energy region where, due to the steep trigger efficiency, the ionization quenching factor degenerates with the rate of detected events. The resulting anti-correlation is shown in the inset in Fig. \ref{fig:trigger_effect}. For data below 1\keVnr, the addition of the constraint via the pull-term allows to reduce the statistical uncertainties by a factor of 1.5 to 2 with respect to an analysis without any rate constraint.

\begin{figure}
	\centering
	\includegraphics[width=0.405\textwidth]{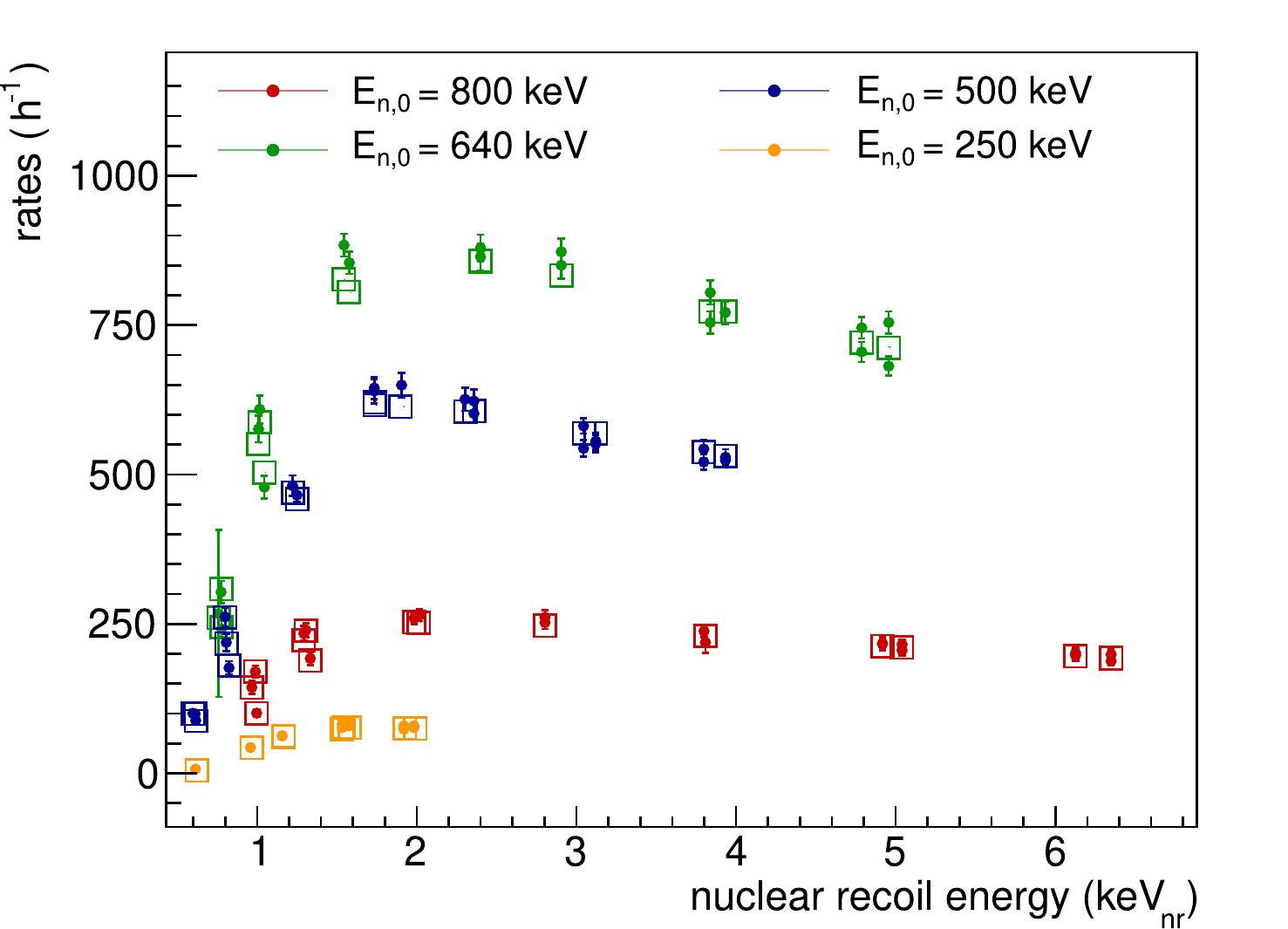}
	\caption{Best fit coincidence rates (points) for each data set and corresponding expectation (square) calculated from the global norm $n^d_0$, the neutron scattering cross section and the trigger efficiency. Rates are given per hour. Differences in current and neutron yield of the Li(p,n) reaction as a function of the energy explain the overall rate difference between the data sets.\label{fig:coincidence_rates}}
\end{figure}

\begin{figure}
	\centering
	\includegraphics[width=0.43\textwidth]{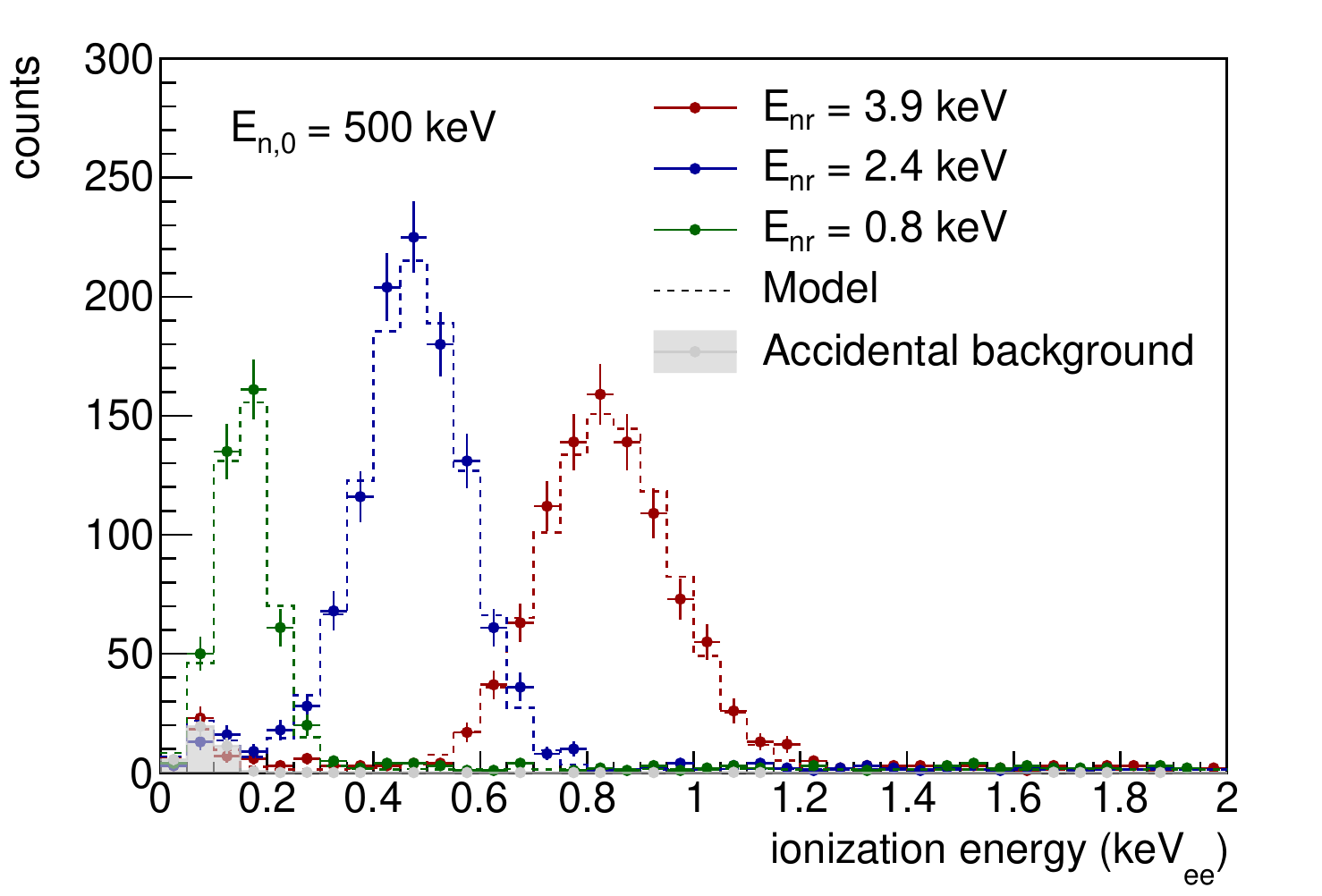}
	\caption{Coincidence energy distributions obtained in the HPGe detector after coincidence selection with three LS detectors at different angles for the data set with the neutron beam energy E$_{n,0}=500\,$keV. The accidental contribution, here in gray, is determined by opening random windows and consists of electronic noise. The best-fit models are superimposed in dashed lines.}\label{fig:500keV_three_positions}
\end{figure}

Best-fit values of the coincidence rate for each quenching data set along with their expected values are shown in Fig. \ref{fig:coincidence_rates}. They are in a very good agreement, confirming a good understanding of all involved effects and systematics. The decreasing trend above 2\keVnr~reflects the angular dependency of the scattering cross-section, whereas the drop for $\lesssim$\,1.5\keVnr~is due to the loss of trigger efficiency and shows therefore the same behavior for all datasets. It becomes also clear that with counting rates of a few hundred counts per hour, the statistics is sufficient for energies above 1-2\keVnr, whereas due to trigger efficiency effects, the available statistics notably drops below 1\keVnr. Fig. \ref{fig:500keV_three_positions} shows coincidence distributions for three exemplary scattering angles at fixed beam energy along with their best-fit model.

Regarding the width of the nuclear recoil distributions, best-fit values of $\sigma_{nr}$ exceed the MC predictions by $(30-60)$\,\%. Possible explanations are an additional smearing of the quenching factor due to the stochastic nature of ion collisions or a mismodeling of the energy spread of the neutron beam. However, the latter case seems to be disfavored: the estimated proton energy losses inside the Li target have been compared with SRIM calculations \cite{srim} and validated by previous measurements at the PIAF facility. Moreover, MC simulations show that the contribution of scattered neutrons at the exit of the collimator is at the percent level, which cannot explain the large discrepancy observed.
For a better quantification of the additional broadening and for a better comparison with other experiments, the following setup-independent quantities have been defined:

\begin{align}
    &\sigma_{nr,\,missing}^2 \equiv \sigma_{nr,\,exp}^2~-~\sigma_{nr,\,MC}^2\\
    &\mathcal{B} \equiv \sigma_{nr,\,missing}^2/E_{nr}^2
\end{align}
where $\sigma_{nr,\,MC}$ is the width predicted from the MC simulation and $\sigma_{nr,\,exp}$ are the ones derived from the data, i.e. the values of $\sigma_{nr,i}$ extracted from the fit of the experimental energy distributions (cf. Eq.~\ref{eq:chi2_pull}).
%
The quantity $\sigma_{nr,\,missing}$ was found to scale roughly with $\sqrt{E_{nr}}$ and amounts to $\sim$\,0.3\keVnr~at 2\keVnr~and 0.55\keVnr~at 5\keVnr. The corresponding values of the broadening factor $\mathcal{B}$ agree with the ones measured in \cite{li2022}, another recent Ge-based setup.

\begin{figure}
	\centering
	\includegraphics[width=0.45\textwidth]{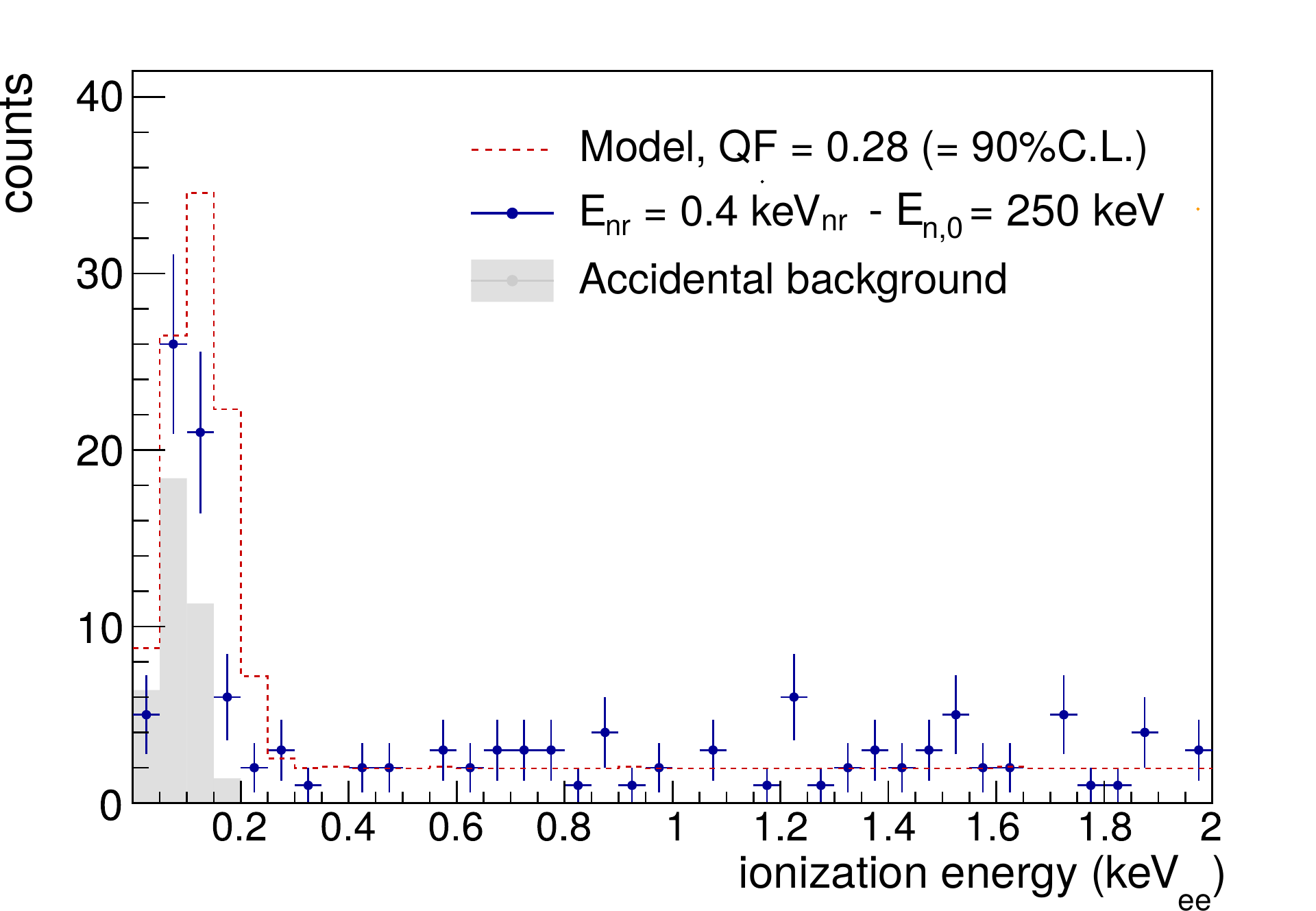}
	\caption{Coincidence energy distribution obtained in the HPGe detector after coincidence selection E$_{nr}$ = 0.4\,\keVnr (blue) along with the estimated accidental contribution (gray). The upper limit at 90\,\% C.L. corresponds to a quenching factor of 0.28 and the corresponding expectation is shown as dashed red line.}\label{fig:upper_limit}
\end{figure}

The coincidence energy distribution corresponding to the lowest nuclear recoil energy was obtained with the lowest beam energy $E_{n,0}$\,=\,250\,keV and corresponds to an expected nuclear recoil energy of 0.4\keVnr. Because of the large statistical uncertainty, a conservative approach was favored and an upper limit in terms of the ionization quenching factor was derived from a likelihood analysis, making use of the normalization constraints from the 250\,keV data set. The upper limit at 90\,\% C.L. was determined from the expected likelihood distributions obtained via a toy MC. The energy distributions of the 0.4\keVnr~coincidence data and the model for the corresponding estimated upper limit for the quenching factor are shown in Fig. \ref{fig:upper_limit}. The same procedure was applied for the coincidence data at 0.6\keVnr~from the same beam energy data set. Both limits are included in Fig. \ref{fig:result}.

\section{Results and discussion} \label{section:results}

\subsection{Combination of data}
 
\begin{figure*}
	\centering
	\includegraphics[width=0.67\textwidth]{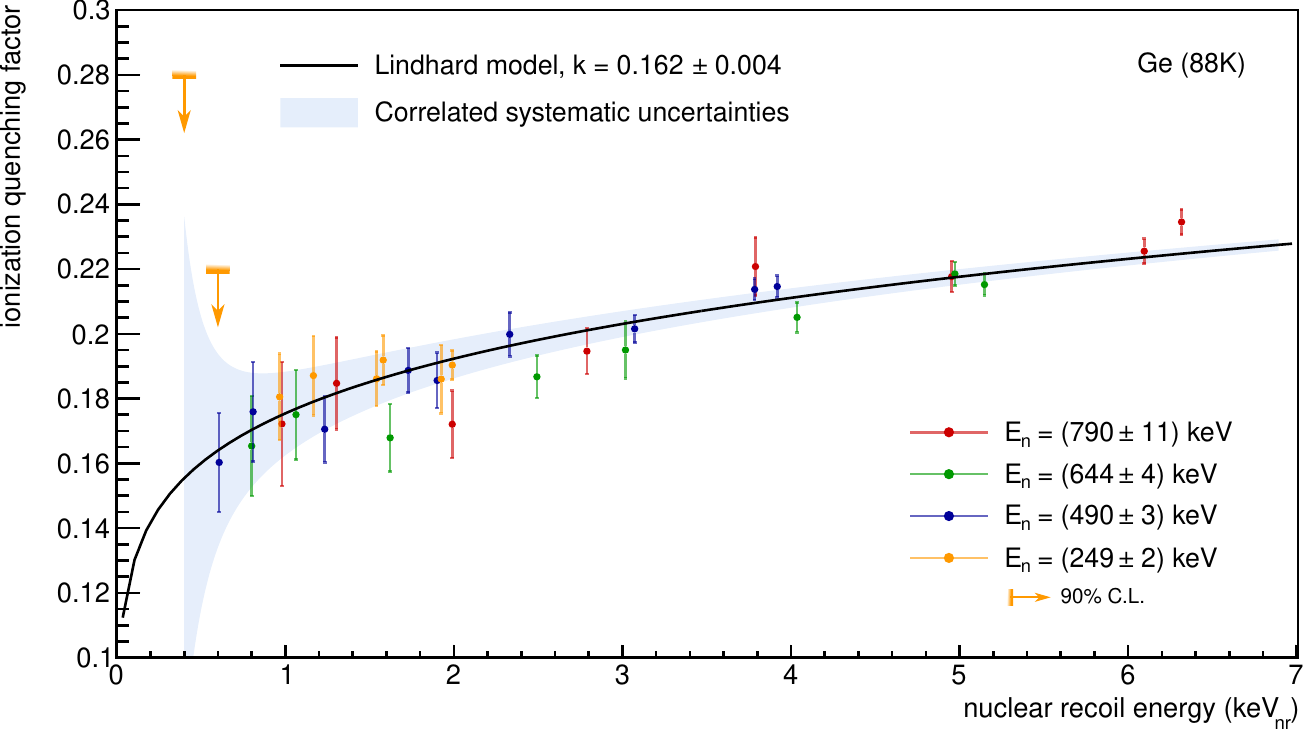}
	\caption{Ionization quenching factor as a function of the nuclear recoil energy. The data points are obtained for the four different data sets, each one corresponding to a different nominal neutron beam energy. Due to the low statistics for the beam energy at 250\,keV only upper limits were extracted for the lowest energy data points. They are represented by the orange arrows. Indicated error bars are uncorrelated uncertainties (statistics and spatial coordinates measurements for each angle), whereas the correlated uncertainties (beam energy, energy scale and trigger efficiency) are represented by the blue band. The best-fit of these data points within a Lindhard theory description is obtained for $k\,=\,0.162\,\pm\,0.004$ (stat+sys) and is illustrated by the black curve.}\label{fig:result}
\end{figure*}

The ionization quenching factor obtained from the fits for each coincidence distribution are provided in Tab. \ref{tab:qf_results_250keV}, \ref{tab:qf_results_500keV}, \ref{tab:qf_results_640keV} and \ref{tab:qf_results_800keV} for the data sets with E$_{n, 0}$ = 250, 500, 640 and 800\,keV, respectively. In the last column the total uncorrelated uncertainties are reported, including both the statistical and the systematic uncertainties from the spatial coordinates measurement. For most of the cases, the latter dominates.
For each beam energy, the same angles were measured several times in each hemisphere and LS detectors were switched in order to cross-check for systematic effects related to detector positioning and neutron detection. No significant discrepancies were found. Therefore, the corresponding data were merged for better visibility and are illustrated in Fig. \ref{fig:result}.

The data points were combined via a fit within the semi-empirical Lindhard theory \cite{Lindhard1963} which describes the energy dependence of the ionization quenching factor $q(E_{nr})$ with a single free parameter $k$:
\begin{equation}
	E_{ee} = q(E_{nr})\cdot E_{nr} = \frac{k\,g(\varepsilon)}{1+k\,g(\varepsilon)}\cdot E_{nr}
\end{equation}
where $E_{nr}$ is the recoil energy in keV and, for germanium, $\varepsilon\,=\,11.5\,Z^{-7/3}E_{nr}$ and $g(\varepsilon)\,=\,3\,\varepsilon^{0.15} + 0.7\,\varepsilon^{0.6} + \varepsilon$. This model allowed for a good description of the data and did not require the inclusion of modified theories with additional parameters, such as the adiabatic correction proposed in \cite{Sorensen2015}.

The complementarity between the different beam energies was exploited to access the same nuclear recoils energies with a different combination of systematic uncertainties.
From Fig. \ref{fig:beam_energy}, it can for instance be seen that the expected width of the coincidence energy distributions strongly varies between the different energies. Moreover, a bias in the spatial measurement of the scattering angles would impact the results obtained from the different beam energies in a different manner. The systematic uncertainty for a given neutron beam energy $E_{n}$ estimated via TOF measurement is fully correlated between the corresponding data points. As reported on Tab. \ref{table:fit_lindhard_results}, I and II, best-fit values of $k$ are consistent for the different beam energies.

Since the fit is driven by the higher energy points with the smallest uncertainties, best-fit values were also compared separately in a restricted low energy range below 1.8\keVnr~and in a high energy range above 1.8\keVnr. Best-fit values of $k$ are consistent for the two energy ranges as reported in Tab. \ref{table:fit_lindhard_results}, I vs. II, for the different data sets.

\begin{table}
	\centering
	\begin{tabular}{l | l l | r r }
	    \hline
		 \multicolumn{2}{l}{data subset} & & $k$ & $\delta k_{tot}$ \\
		\hline
		 & runs 3, 4 & $E_{n,0}$ = 250\,keV & 0.157 & 0.007 \\
		 & runs 7, 8, 9 & $E_{n,0}$ = 500\,keV & 0.163 & 0.005 \\
		I & runs 5, 6 & $E_{n,0}$ = 640\,keV & 0.157 & 0.004 \\
		 & runs 10, 11 & $E_{n,0}$ = 800\,keV & 0.162 & 0.005\\
		\hdashline
		 & \multicolumn{2}{l}{combined $[1.8,\,6.3]$\keVnr} & 0.162 & 0.004\\
		\hline
		 & runs 3, 4 & $E_{n,0}$ = 250\,keV & 0.167 & 0.011\\
		II & runs 7, 8, 9 & $E_{n,0}$ = 500\,keV & 0.157 & 0.012\\
		 & runs 5, 6 & $E_{n,0}$ = 640\,keV & 0.148 & 0.013\\
		 & runs 10, 11 & $E_{n,0}$ = 800\,keV & 0.169 & 0.017\\
		\hdashline
		 & \multicolumn{2}{l}{combined $[0.6,\,1.8]$\keVnr} & 0.161 & 0.011\\
		\hline
		    & pixel-1 & & 0.159 & 0.004\\
		III & pixel-2 & & 0.162 & 0.004\\
		    & pixel-3 & & 0.160 & 0.004\\
		    & pixel-4 & & 0.162 & 0.004\\
		\hline
		\multicolumn{3}{l}{combined all} & 0.162 & 0.004\\
		\hline
	\end{tabular}
	\caption{Best-fit of quenching data separated into different data sets (divided into beam energies and pixels).}\label{table:fit_lindhard_results}
\end{table}

The uncertainties related to the HPGe detector (energy scale, detector response) are treated as fully correlated between all data points. 
The systematic uncertainty on the energy scale was derived by allowing for a constant shift $\Delta E_{ee} = \pm\,12$\,eV$_{ee}$ in the reconstructed ionization energy, implying a larger effect towards lower energies. The value comes from the constraint on the energy scale obtained with the 1.30\keVee~L-shell atomic deexcitation line following Ge EC decays, as discussed in Sect. \ref{subsection:hpge}.
In addition, to quantify the impact of a potential mismodeling of the detector response close to the threshold, the mean position of the trigger efficiency curve was shifted by $\pm$\,10\,eV$_{ee}$ which is a conservative upper limit for the four pixels. The impact is negligible above 1.2\keVnr.
Due to the anti-correlation between detected rate and ionization quenching factor (discussed in Sect.  \ref{subsection:analysis}), the impact grows rapidly below this energy. It becomes the dominant contribution for the lowest energy points at $E_{nr} \lesssim 0.8$\keVnr, as illustrated by the enlarged uncertainty band in Fig. \ref{fig:result}. 

Advantage was taken of the pixelized structure of the HPGe detector to validate these estimations. Indeed, as they were individually calibrated and characterized, the four pixels can be considered as almost independent detectors. Moreover, different acquisition conditions in terms of noise rate were used on purpose such that the accidental contribution varies by about one order of magnitude between the pixels. The full analysis was then repeated for each pixel independently. The results (reported in Tab. \ref{table:fit_lindhard_results}, III) are compatible with the results under I and II and do not indicate any underestimated source of uncertainty.

Tab. \ref{tab:error_budget} summarizes all the above mentioned sources of systematic uncertainties and their contributions to the final results. For $E_{nr} \gtrsim$ 2\keVnr, the statistical uncertainty is negligible and the correlated uncertainty on the energy scale of the HPGe detector is the major contributor to the total quoted uncertainty, followed by the systematic uncertainty on the beam energy and of the geometry of the setup. In the low energy range, statistical uncertainties are not negligible anymore. The correlated uncertainty on the energy scale is still the dominant contribution. Furthermore, as emphasized by the enlarged uncertainty band in Fig. \ref{fig:result}, the uncertainty due to the modeling of the trigger efficiency dominates the sub-keV region.
The large contribution of the uncertainty of the energy scale is related to the lack of photon sources in the sub-\keVee~region which could be used for calibration. A smaller statistical uncertainty from the 1.30\keVee~activation line would have allowed to put a more stringent constraint on the energy scale. The present approach is therefore conservative and does also cover the uncertainties related to the small non-linearities observed in the few hundreds of eV$_{ee}$ region, as illustrated by the gray band in the bottom panel of Fig.\,\ref{fig:trigger_efficiency}.
For the sub-\keVnr~region, the precision of the measurement is intrinsically limited by the trigger efficiency of the HPGe detector. A precise knowledge of the detector response at these energies allows partly to overcome this limitation but HPGe detectors with thresholds below 100\,eV$_{ee}$ are needed to access the sub-\keVnr~region with a precision better than $\sim$\,0.01 on the ionization quenching factor.

To summarize, the combined fit of all data points between 0.6\keVnr~and 6.3\keVnr~using the Lindhard theory yields a best fit value of $k\,=\,0.162$ with a total uncertainty (stat+syst) of 0.004. It is represented by the black line in Fig. \ref{fig:result}. 

\begin{table*}
	\centering
	\begin{tabular*}{\textwidth}{l | c c c | l l | c c }
	     \multicolumn{6}{c}{} & \multicolumn{2}{c}{uncertainty on $k$}\\
		source & point & data set & all & constraint & value & $[2,6.3]$\keVnr & $[0.6, 2]$\keVnr\\
		\hline
		statistics & $\times$ & & & -- & -- & negligible & 0.001\\
		geometry & $\times$ & & & spatial coord. meas. & (0.5-1.5)$^o$ & 0.001 & 0.002\\
		beam energy & & $\times$ & & TOF measurements & $\sim$1\,\% (cf. Tab. \ref{tab:runs_tof_summary}) & 0.001 & 0.001\\
		energy scale & & & $\times$ & Fe-55 + Ge related lines & 12\,eV$_{ee}$ & 0.003 & 0.008\\
		trigger efficiency & & & $\times$ &pulser measurements & 10\,eV$_{ee}$ & negligible & 0.001\\
		\hline
		total & & & & & & 0.003 & 0.008
	\end{tabular*}
	\caption{Summary of the uncertainties sources taken into account in the analysis. Statistics and geometry uncertainties are uncorrelated between the points, whereas beam energy uncertainties are correlated for each data set. HPGe detector modeling effects are fully correlated. Their impact in terms of absolute uncertainty on the quenching parameter $k$ of the Lindhard theory is given in the two last columns, for two different energy ranges.}\label{tab:error_budget}
\end{table*}

\subsection{Analysis of the integral energy distribution of recoiling nuclei}\label{subsection:integrated_recoils}

In this last section an additional cross-check is proposed, considered as almost independent of the main result of this article. A complementary approach to determine ionization quenching factors consists in comparing the energy distribution of recoil nuclei integrated over all scattering angles (i.e. without coincidence requirements) with a MC simulation of the experiment. This approach was for instance used in \cite{Scholz2016, Chavarria2016, Collar2021}. Such an analysis relies on an accurate modeling of the setup and a quenching model. Its parameters can be tuned in the simulation in order to reproduce the observed data.
Although we strongly favor the direct and model-independent technique of a coincidence measurement, advantage was taken of the high-statistics neutron recoil data integrated over all angles and for different neutron beam energies in the HPGe detector collected as a by-product during data collection. Note that the range of recoil energies probed in this way is much broader than the one studied by selecting only small scattering angles: from Eq. (\ref{eq:enr_f_theta}), the maximum recoil energy obtained for back-scattered neutrons equals e.g. 26\keVnr~for E$_{n,0}$ = 500\,keV.
Integrated energy distributions of recoil nuclei without coincidence requirements were compared to a \texttt{Geant4}\cite{geant4} simulation of the setup including the detailed geometry of the detector end-cap. A simplified description of the neutron beam was implemented: the beam profile was inferred from TOF measurements at different distances and the energy profile was taken from calculations accounting for the thickness of the lithium \cite{NeuSDesc}.
As for the main analysis, the expected nuclear recoil distribution was convoluted with the detector response $\mathcal{M}_{ee}$ as detailed in Sect. \ref{subsection:hpge}. Since this analysis is intended to be a cross-check of the main coincidence analysis, only the ionization quenching factor description within the Lindhard theory with the best-fit k\,=\,0.162 was implemented.

\begin{figure}
	\centering
	\includegraphics[width=0.42\textwidth]{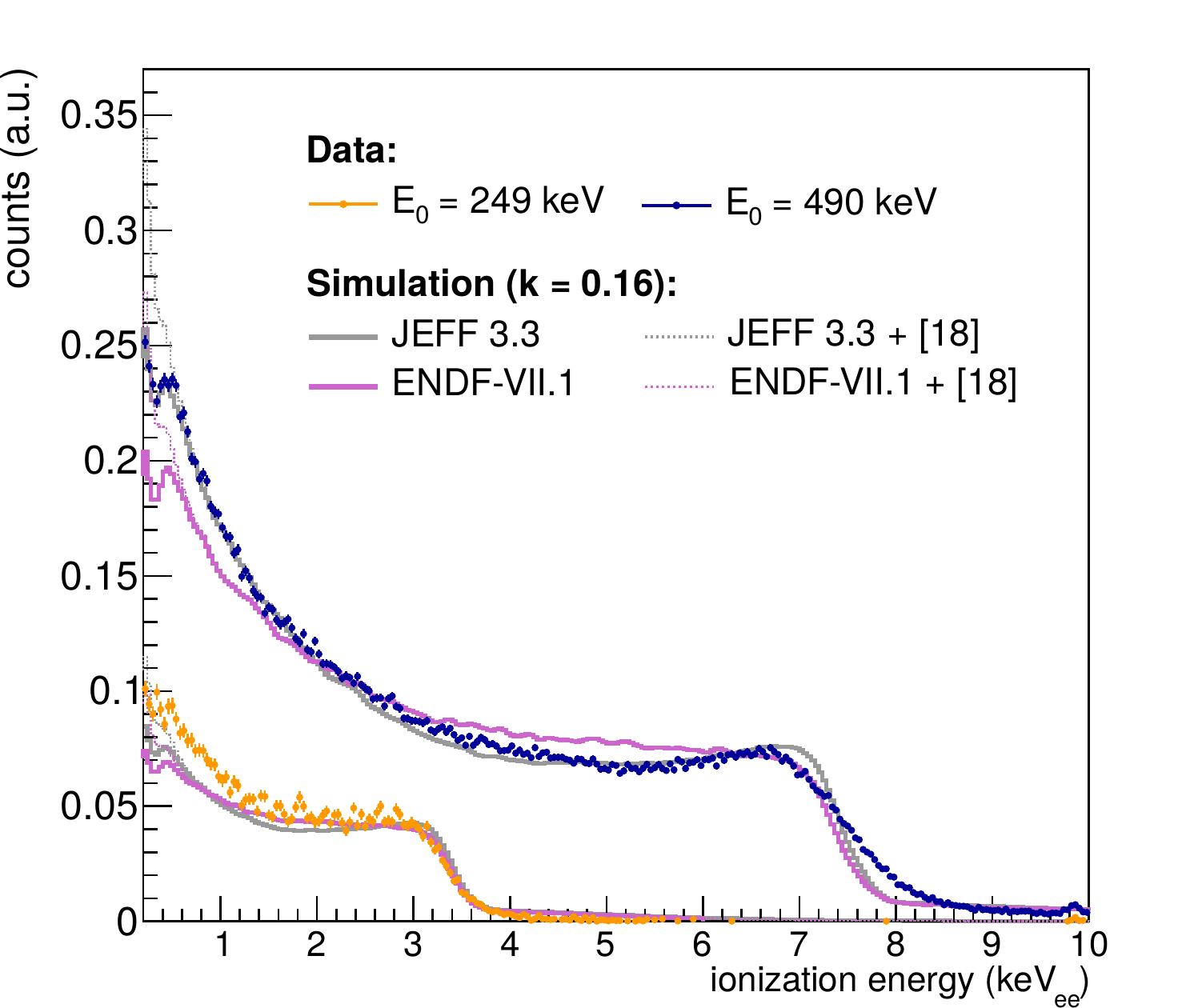}
	\caption{Measured integrated nuclear recoil distribution for E$_{n}$ = 249\,keV (orange points) and E$_{n}$ = 490\,keV (blue points) and associated \texttt{Geant4} simulations (solid lines) with JEFF~3.3 (gray) and ENDF-BVII.1 (purple). Superimposed are the simulations including the extension of the quenching factor model found in \cite{Collar2021} (dashed lines).}\label{fig:integrated_spectra}
\end{figure}

An overall reasonable agreement was found for the four data sets, especially concerning the position of the end-point in energy, as shown in Fig. \ref{fig:integrated_spectra}. Detector resolution effects -- described by Eq. (\ref{eq:energy_resolution}) and the constraint of the Fe-55 calibration at these energies -- cannot explain the steeper edge of the end-point found in the simulation. An additional energy broadening resulting from the collimator could be the origin of the discrepancy. The impact of surrounding materials and of the beam profile were investigated and found to play only a secondary role for the spectral shape of the energy distribution. However, it is worth noting that the shape of the energy distribution strongly depends on the evaluated nuclear database. In particular, a much better agreement was found for the region below 1\keVnr~when making use of JEFF 3.3\cite{jeff33} instead of ENDF-BVII.1\cite{endfb7}. This dependence on the databases in the MC simulation illustrates the difficulty of extracting quenching factors out of integrated distributions and it reinforces the need of kinematically constrained measurements like the one presented in this work.

A steep increase of the ionization quenching factor below 1\keVnr~as measured in \cite{Collar2021} was also implemented. The resulting energy distributions of recoil nuclei for the two evaluated nuclear databases are shown by the dashed lines in Fig. \ref{fig:integrated_spectra}. The addition of such an enhancement only affects the lowest part of the energy distribution and does not significantly improve the overall description of the data.

\subsection{Conclusion}

A direct and precise measurement of the ionization quenching factor for nuclear recoils in germanium was presented. A HPGe detector with a mass of 10\,g and a thickness of 6\,mm was operated at 88\,K and exposed to monoenergetic neutron beams with energies ranging from 250\,keV to 800\,keV at the accelerator facility PIAF of PTB in Braunschweig (Germany). Nuclear recoil energies between 0.4\keVnr~and 6.3\keVnr~were selected by the detection in coincidence of the scattered neutrons in LS detectors.
In this energy range, data are compatible with the Lindhard theory prediction with $k\,=\,0.162\,\pm\,0.004$ (stat+syst), as presented in Fig. \ref{fig:result}.

The uncertainties of the measurement were discussed in detail. Their contributions are summarized in Tab. \ref{tab:error_budget}. Special attention was paid to constrain them by performing multiple cross-checks, e.g. by using different monoenergetic neutron beams, by interchanging the neutron coincidence detectors and by taking advantage of the segmented structure of the target detector.
In the region of interest of the current experiments looking for \CEvNS~at nuclear power reactors with expected nuclear recoil energies above 1.5\,\keVnr, a measurement of the quenching factor with a relative uncertainty of a few percent is provided. This was achieved by determining the scattering angles with uncertainties of about one degree and by precisely monitoring the energy of the neutron beam.
For recoil energies of less than $\sim$\,1\keVnr, the fully correlated systematic uncertainties related to HPGe detector response modeling dominate.
In particular, energy threshold effects -- and therefore decreasing statistics -- illustrated the difficulty of extracting ionization quenching factors at these energies even with improved state-of-the-art HPGe detectors.
Thanks to a precise characterization of the detector response and a very good control on the coincidence rates, this limitation could partly be overcome and quenching factors were extracted with a precision of (5-10)\,\% from data in this low energy region.

As an additional independent cross-check, a comparison of the integrated recoil energy spectrum (i.e. collected in the HPGe without coincidence requirement) with a MC simulation of the setup was performed and an overall good agreement was found when making use of the JEFF 3.3 evaluated nuclear database.

This work significantly contributes to the understanding of the ionization quenching factor of keV nuclear recoils in Ge. Above 1\keVnr, the quenching factor was measured with a precision of a few percent. The data follow the Lindhard theory and are in agreement with several previous measurements that were performed at liquid nitrogen temperature in the 0.5-10\keVnr~range \cite{Jones1971, Jones1975, Messous1995, Barbeau2007}. However, this result is found to fully disagree with a recent precision measurement at 50\,mK \cite{cdms2022} in which a significant drop of the quenching factor below 7\keVnr~was found. In this case the measurement technique and experimental conditions (temperature, electric field) were nevertheless significantly different from the ones reported in this article. Below 1\keVnr, our data do not confirm the outcome of another recent measurement in Ge at 77\,K that suggests an enhanced quenching factor with respect to the Lindhard prediction \cite{Collar2021}.
%
Our analysis does include a refined description of non-linearities and trigger efficiency effects affecting this low energy regime, refuting in particular the claims raised in \cite{collar_comment} and the proposed correction to obtain a better agreement with the result found in \cite{Collar2021}.
However, resolving these discrepancies is essential for the present and future experiments based on HPGe detectors in the search for coherent elastic-neutrino nucleus scattering and light dark matter.


\small
\textbf{Acknowledgements}
We thank all involved divisions and workshops at the Max-Planck-Institut für Kernphysik in Heidelberg, in particular R. Kankanyan for the design, R. Heldner for helping us mounting the experiment and K. Zink and S. Kolb for their IT support.
Further thanks are directed to Dr. S. Schoppmann and B. Gramlich (MPIK) for their contribution on the liquid scintillators.
We express our gratitude to Dr. Andreas Zimbal and the PTB staff involved in our measurement campaign for their great technical, mechanical and logistical support, with a particular mention to Stefan Loeb and Benedikt Schulze. We thank Mirion (Canberra) Technologies in Lingolsheim (France) for their prolific support. This measurement is supported financially by the Max Planck Society (MPG), T. Rink by the German Research Foundation (DFG) through the research training group “Particle physics beyond the Standard Model” (GRK 1940), and together with J. Hakenmüller by the International Max Planck Research School for Precision Tests of Fundamental Symmetries.
\normalsize


\bibliographystyle{bibliostyle}
\bibliography{literature}

\begin{thebibliography}{10}
\expandafter\ifx\csname urlstyle\endcsname\relax
  \providecommand{\doi}[1]{doi:\discretionary{}{}{}#1}
  \providecommand{\eprint}[1]{arXiv:\discretionary{}{}{}#1}\else
  \providecommand{\doi}{doi:\discretionary{}{}{}\begingroup \urlstyle{rm}\Url}
  \providecommand{\eprint}{arXiv:\discretionary{}{}{}\begingroup
  \urlstyle{rm}\Url}\fi

\bibitem{Sattler1966}
A.~R. Sattler, F.~L. Vook and J.~M. Palms.
\newblock \emph{Ionization Produced by Energetic Germanium Atoms within a
  Germanium Lattice}.
\newblock Phys. Rev., \textbf{143}:588--594, 1966.
\newblock
  \href{http://dx.doi.org/10.1103/PhysRev.143.588}{\doi{10.1103/PhysRev.143.588}}.

\bibitem{Chasman1967}
C.~Chasman, K.~W. Jones, R.~A. Ristinen and J.~T. Sample.
\newblock \emph{{Measurement of the Energy Loss of Germanium Atoms to Electrons
  in Germanium at Energies below 100 keV. II}}.
\newblock Phys. Rev., \textbf{154}:239--244, 1967.
\newblock
  \href{http://dx.doi.org/10.1103/PhysRev.154.239}{\doi{10.1103/PhysRev.154.239}}.

\bibitem{Shutt1992}
T.~Shutt et~al.
\newblock \emph{Measurement of ionization and phonon production by nuclear
  recoils in a 60 g crystal of germanium at 25 mK}.
\newblock Phys. Rev. Lett., \textbf{69}:3425--3427, 1992.
\newblock
  \href{http://dx.doi.org/10.1103/PhysRevLett.69.3425}{\doi{10.1103/PhysRevLett.69.3425}}.

\bibitem{Baudis1998}
L.~Baudis et~al.
\newblock \emph{{High purity germanium detector ionization pulse shapes of
  nuclear recoils, gamma interactions and microphonism}}.
\newblock Nucl. Instrum. Meth. A, \textbf{418}:348--354, 1998.
\newblock
  \href{http://dx.doi.org/10.1016/S0168-9002(98)00782-7}{\doi{10.1016/S0168-9002(98)00782-7}}.
\newblock \href{https://arxiv.org/abs/hep-ex/9901028}{\eprint{hep-ex/9901028}}.

\bibitem{Simon2003}
E.~Simon et~al.
\newblock \emph{SICANE: a detector array for the measurement of nuclear recoil
  quenching factors using a monoenergetic neutron beam}.
\newblock Nucl. Instrum. Meth. A, \textbf{507}(3):643--656, 2003.
\newblock ISSN 0168-9002.
\newblock
  \href{http://dx.doi.org/10.1016/S0168-9002(03)01438-4}{\doi{10.1016/S0168-9002(03)01438-4}}.

\bibitem{Benoit2007}
A.~Benoit et~al.
\newblock \emph{{Measurement of the response of heat-and-ionization germanium
  detectors to nuclear recoils}}.
\newblock Nucl. Instrum. Meth. A, \textbf{577}:558--568, 2007.
\newblock
  \href{http://dx.doi.org/10.1016/j.nima.2007.04.118}{\doi{10.1016/j.nima.2007.04.118}}.
\newblock
  \href{https://arxiv.org/abs/astro-ph/0607502}{\eprint{astro-ph/0607502}}.

\bibitem{Lindhard1963}
J.~Lindhard.
\newblock \emph{Influence of Crystal Lattice on Motion of Energetic Charged
  Particles}.
\newblock Kongel. Dan. Vidensk. Selsk., Mat.-Fys. Medd., \textbf{34}(14), 1965.

\bibitem{Bonet2021d}
H.~Bonet et~al.
\newblock \emph{{Large-size sub-keV sensitive germanium detectors for the CONUS
  experiment}}.
\newblock Eur. Phys. J. C, \textbf{81}(3):267, 2021.
\newblock
  \href{http://dx.doi.org/10.1140/epjc/s10052-021-09038-3}{\doi{10.1140/epjc/s10052-021-09038-3}}.
\newblock \href{https://arxiv.org/abs/2010.11241}{\eprint{2010.11241}}.

\bibitem{Soma2014}
A.~K. Soma et~al.
\newblock \emph{{Characterization and Performance of Germanium Detectors with
  sub-keV Sensitivities for Neutrino and Dark Matter Experiments}}.
\newblock Nucl. Instrum. Meth. A, \textbf{836}:67--82, 2016.
\newblock
  \href{http://dx.doi.org/10.1016/j.nima.2016.08.044}{\doi{10.1016/j.nima.2016.08.044}}.
\newblock \href{https://arxiv.org/abs/1411.4802}{\eprint{1411.4802}}.

\bibitem{Belov2015}
V.~Belov et~al.
\newblock \emph{The $\upnu${GeN} experiment at the Kalinin Nuclear Power
  Plant}.
\newblock Journal of Instrumentation, \textbf{10}(12):P12011--P12011, 2015.
\newblock
  \href{http://dx.doi.org/10.1088/1748-0221/10/12/p12011}{\doi{10.1088/1748-0221/10/12/p12011}}.

\bibitem{Collar2021b}
J.~Colaresi et~al.
\newblock \emph{First results from a search for coherent elastic
  neutrino-nucleus scattering at a reactor site}.
\newblock Phys. Rev. D, \textbf{104}:072003, 2021.
\newblock
  \href{http://dx.doi.org/10.1103/PhysRevD.104.072003}{\doi{10.1103/PhysRevD.104.072003}}.

\bibitem{Sorensen2015}
P.~Sorensen.
\newblock \emph{Atomic limits in the search for galactic dark matter}.
\newblock Phys. Rev. D, \textbf{91}:083509, 2015.
\newblock
  \href{http://dx.doi.org/10.1103/PhysRevD.91.083509}{\doi{10.1103/PhysRevD.91.083509}}.

\bibitem{Sarkis2020}
Y.~Sarkis, A.~Aguilar-Arevalo and J.~C. D'Olivo.
\newblock \emph{Study of the ionization efficiency for nuclear recoils in pure
  crystals}.
\newblock Phys. Rev. D, \textbf{101}:102001, 2020.
\newblock
  \href{http://dx.doi.org/10.1103/PhysRevD.101.102001}{\doi{10.1103/PhysRevD.101.102001}}.

\bibitem{Sarkis2021}
Y.~Sarkis, A.~Aguilar-Arevalo and J.~C. D'Olivo.
\newblock \emph{A Study of the Ionization Efficiency for Nuclear Recoils in
  Pure Crystals}.
\newblock Phys. Atom. Nuclei, \textbf{84}:590–594, 2021.
\newblock
  \href{http://dx.doi.org/10.1134/S1063778821040268}{\doi{10.1134/S1063778821040268}}.

\bibitem{Barker2012}
D.~Barker and D.-M. Mei.
\newblock \emph{Germanium detector response to nuclear recoils in searching for
  dark matter}.
\newblock Astroparticle Physics, \textbf{38}:1--6, 2012.
\newblock ISSN 0927-6505.
\newblock
  \href{http://dx.doi.org/10.1016/j.astropartphys.2012.08.006}{\doi{10.1016/j.astropartphys.2012.08.006}}.

\bibitem{Chavarria2016}
A.~E. Chavarria et~al.
\newblock \emph{Measurement of the ionization produced by sub-keV silicon
  nuclear recoils in a CCD dark matter detector}.
\newblock Phys. Rev. D, \textbf{94}:082007, 2016.
\newblock
  \href{http://dx.doi.org/10.1103/PhysRevD.94.082007}{\doi{10.1103/PhysRevD.94.082007}}.

\bibitem{Izraelevitch2017}
F.~Izraelevitch et~al.
\newblock \emph{A measurement of the ionization efficiency of nuclear recoils
  in silicon}.
\newblock Journal of Instrumentation, \textbf{12}(06):P06014--P06014, 2017.
\newblock
  \href{http://dx.doi.org/10.1088/1748-0221/12/06/p06014}{\doi{10.1088/1748-0221/12/06/p06014}}.

\bibitem{Collar2021}
J.~I. Collar, A.~R.~L. Kavner and C.~M. Lewis.
\newblock \emph{Germanium response to sub-keV nuclear recoils: A multipronged
  experimental characterization}.
\newblock Phys. Rev. D, \textbf{103}:122003, 2021.
\newblock
  \href{http://dx.doi.org/10.1103/PhysRevD.103.122003}{\doi{10.1103/PhysRevD.103.122003}}.

\bibitem{cdms2022}
M.~F. Albakry et~al.
\newblock \emph{Ionization yield measurement in a germanium CDMSlite detector
  using photo-neutron sources}.
\newblock Phys. Rev. D, \textbf{105}:122002, 2022.
\newblock
  \href{http://dx.doi.org/10.1103/PhysRevD.105.122002}{\doi{10.1103/PhysRevD.105.122002}}.

\bibitem{Jones1971}
K.~W. Jones and H.~W. Kraner.
\newblock \emph{{Stopping of 1- to 1.8-keV $^{73}$Ge Atoms in Germanium}}.
\newblock Phys. Rev. C, \textbf{4}:125--129, 1971.
\newblock
  \href{http://dx.doi.org/10.1103/PhysRevC.4.125}{\doi{10.1103/PhysRevC.4.125}}.

\bibitem{Jones1975}
K.~W. Jones and H.~W. Kraner.
\newblock \emph{{Energy lost to ionization by 254-eV $^{73}$Ge atoms stopping
  in Ge}}.
\newblock Phys. Rev. A, \textbf{11}:1347--1353, 1975.
\newblock
  \href{http://dx.doi.org/10.1103/PhysRevA.11.1347}{\doi{10.1103/PhysRevA.11.1347}}.

\bibitem{Messous1995}
Y.~Messous.
\newblock \emph{{Calibration of a Ge crystal with nuclear recoils for the
  development of a dark matter detector}}.
\newblock Astropart. Phys., \textbf{3}:361--366, 1995.
\newblock
  \href{http://dx.doi.org/10.1016/0927-6505(95)00007-4}{\doi{10.1016/0927-6505(95)00007-4}}.

\bibitem{Barbeau2007}
P.~S. Barbeau, J.~I. Collar and O.~Tench.
\newblock \emph{Large-mass ultralow noise germanium detectors: performance and
  applications in neutrino and astroparticle physics}.
\newblock Journal of Cosmology and Astroparticle Physics,
  \textbf{2007}(09):009--009, 2007.
\newblock
  \href{http://dx.doi.org/10.1088/1475-7516/2007/09/009}{\doi{10.1088/1475-7516/2007/09/009}}.

\bibitem{Scholz2016}
B.~J. Scholz, A.~E. Chavarria, J.~I. Collar, P.~Privitera and A.~E. Robinson.
\newblock \emph{{Measurement of the low-energy quenching factor in germanium
  using an $^{88}$Y/Be photoneutron source}}.
\newblock Phys. Rev. D, \textbf{94}(12):122003, 2016.
\newblock
  \href{http://dx.doi.org/10.1103/PhysRevD.94.122003}{\doi{10.1103/PhysRevD.94.122003}}.
\newblock \href{https://arxiv.org/abs/1608.03588}{\eprint{1608.03588}}.

\bibitem{Bonet2021}
H.~Bonet et~al.
\newblock \emph{{Constraints on Elastic Neutrino Nucleus Scattering in the
  Fully Coherent Regime from the CONUS Experiment}}.
\newblock Phys. Rev. Lett., \textbf{126}(4):041804, 2021.
\newblock
  \href{http://dx.doi.org/10.1103/PhysRevLett.126.041804}{\doi{10.1103/PhysRevLett.126.041804}}.
\newblock \href{https://arxiv.org/abs/2011.00210}{\eprint{2011.00210}}.

\bibitem{Bonet2021b}
H.~Bonet et~al.
\newblock \emph{{Novel constraints on neutrino physics beyond the standard
  model from the CONUS experiment}}.
\newblock JHEP, \textbf{05}:085, 2022.
\newblock
  \href{http://dx.doi.org/10.1007/JHEP05(2022)085}{\doi{10.1007/JHEP05(2022)085}}.
\newblock \href{https://arxiv.org/abs/2110.02174}{\eprint{2110.02174}}.

\bibitem{Brede1980}
H.~Brede et~al.
\newblock \emph{The Braunschweig accelerator facility for fast neutron
  research: 1: Building design and accelerators}.
\newblock Nuclear Instruments and Methods, \textbf{169}(3):349--358, 1980.
\newblock ISSN 0029-554X.
\newblock
  \href{http://dx.doi.org/https://doi.org/10.1016/0029-554X(80)90928-3}{\doi{https://doi.org/10.1016/0029-554X(80)90928-3}}.

\bibitem{mcnp}
C.~J. Werner et~al.
\newblock \emph{MCNP6.2 Release Notes}.
\newblock Los Alamos National Laboratory Report, (LA-UR-18-20808), 2018.

\bibitem{Stuck1973}
R.~Stuck, J.~P. Ponpon, R.~Berger and P.~Siffert.
\newblock \emph{Temperature dependence of the average energy per electron-hole
  pair in Silicon and Germanium}.
\newblock Radiation Effects, \textbf{20}(1-2):75--80, 1973.
\newblock
  \href{http://dx.doi.org/10.1080/00337577308232269}{\doi{10.1080/00337577308232269}}.
\newblock
  \href{https://arxiv.org/abs/https://doi.org/10.1080/00337577308232269}{\eprint{https://doi.org/10.1080/00337577308232269}}.

\bibitem{Pehl1979}
R.~H. Pehl et~al.
\newblock \emph{Radiation Damage Resistance of Reverse Electrode Ge Coaxial
  Detectors}.
\newblock IEEE Transactions on Nuclear Science, \textbf{26}(1):321--323, 1979.
\newblock
  \href{http://dx.doi.org/10.1109/TNS.1979.4329652}{\doi{10.1109/TNS.1979.4329652}}.

\bibitem{Raudorf1987}
T.~W. Raudorf and R.~H. Pehl.
\newblock \emph{Effect of charge carrier trapping on germanium coaxial detector
  line shapes}.
\newblock Nucl. Instrum. Meth. A, \textbf{255}(3):538--551, 1987.
\newblock ISSN 0168-9002.
\newblock
  \href{http://dx.doi.org/10.1016/0168-9002(87)91225-3}{\doi{10.1016/0168-9002(87)91225-3}}.

\bibitem{Descovich2005}
M.~Descovich et~al.
\newblock \emph{Effects of neutron damage on the performance of large volume
  segmented germanium detectors}.
\newblock Nucl. Instrum. Meth. A, \textbf{545}(1):199--209, 2005.
\newblock ISSN 0168-9002.
\newblock
  \href{http://dx.doi.org/10.1016/j.nima.2005.01.308}{\doi{10.1016/j.nima.2005.01.308}}.

\bibitem{Borrel1999}
V.~Borrel et~al.
\newblock \emph{Fast neutron-induced damage in INTEGRAL n-type HPGe detectors}.
\newblock Nucl. Instrum. Meth. A, \textbf{430}(2):348--362, 1999.
\newblock ISSN 0168-9002.
\newblock
  \href{http://dx.doi.org/10.1016/S0168-9002(99)00218-1}{\doi{10.1016/S0168-9002(99)00218-1}}.

\bibitem{Ross2009}
T.~Ross et~al.
\newblock \emph{Neutron damage tests of a highly segmented germanium crystal}.
\newblock Nucl. Instrum. Meth. A, \textbf{606}(3):533--544, 2009.
\newblock ISSN 0168-9002.
\newblock
  \href{http://dx.doi.org/10.1016/j.nima.2009.04.024}{\doi{10.1016/j.nima.2009.04.024}}.

\bibitem{Soma2016}
A.~Soma et~al.
\newblock \emph{Characterization and performance of germanium detectors with
  sub-keV sensitivities for neutrino and dark matter experiments}.
\newblock Nuclear Instruments and Methods in Physics Research Section A:
  Accelerators, Spectrometers, Detectors and Associated Equipment,
  \textbf{836}:67--82, 2016.
\newblock ISSN 0168-9002.
\newblock
  \href{http://dx.doi.org/https://doi.org/10.1016/j.nima.2016.08.044}{\doi{https://doi.org/10.1016/j.nima.2016.08.044}}.

\bibitem{Batchelor1961}
R.~Batchelor, W.~Gilboy, J.~Parker and J.~Towle.
\newblock \emph{The response of organic scintillators to fast neutrons}.
\newblock Nuclear Instruments and Methods, \textbf{13}:70--82, 1961.
\newblock ISSN 0029-554X.
\newblock
  \href{http://dx.doi.org/https://doi.org/10.1016/0029-554X(61)90171-9}{\doi{https://doi.org/10.1016/0029-554X(61)90171-9}}.

\bibitem{Pangher1966}
J.~De~Pangher and L.~L. Nichols.
\newblock \emph{Precision long counter for measuring fast neutron flux
  density}.
\newblock Tech. Rep. BNWL, \textbf{260}, 1966.
\newblock \href{http://dx.doi.org/10.2172/4513481}{\doi{10.2172/4513481}}.

\bibitem{Liskien1975}
H.~Liskien and A.~Paulsen.
\newblock \emph{Neutron production cross sections and energies for the
  reactions 7Li(p,n)7Be and 7Li(p,n)7Be$^*$}.
\newblock Atomic Data and Nuclear Data Tables, \textbf{15}(1):57--84, 1975.
\newblock ISSN 0092-640X.
\newblock
  \href{http://dx.doi.org/https://doi.org/10.1016/0092-640X(75)90004-2}{\doi{https://doi.org/10.1016/0092-640X(75)90004-2}}.

\bibitem{NeuSDesc}
{E. Birgersson and G. L\"{o}vestam}.
\newblock \emph{NeuSDesc-Neutron Source Description Software Manual}.
\newblock EUR 23794 EN. Luxembourg (Luxembourg): OPOCE. JRC51437, 2009.

\bibitem{endfb7}
M.~Chadwick et~al.
\newblock \emph{ENDF/B-VII.1 Nuclear Data for Science and Technology: Cross
  Sections, Covariances, Fission Product Yields and Decay Data}.
\newblock Nuclear Data Sheets, \textbf{112}(12):2887--2996, 2011.
\newblock ISSN 0090-3752.
\newblock
  \href{http://dx.doi.org/10.1016/j.nds.2011.11.002}{\doi{10.1016/j.nds.2011.11.002}}.

\bibitem{jeff33}
A.~J.~M. Plompen et~al.
\newblock \emph{{The joint evaluated fission and fusion nuclear data library,
  JEFF-3.3}}.
\newblock Eur. Phys. J. A, \textbf{56}(7):181, 2020.
\newblock
  \href{http://dx.doi.org/10.1140/epja/s10050-020-00141-9}{\doi{10.1140/epja/s10050-020-00141-9}}.

\bibitem{jendl}
K.~Shibata et~al.
\newblock \emph{JENDL-4.0: A New Library for Nuclear Science and Engineering}.
\newblock Journal of Nuclear Science and Technology, \textbf{48}(1):1--30,
  2011.
\newblock
  \href{http://dx.doi.org/10.1080/18811248.2011.9711675}{\doi{10.1080/18811248.2011.9711675}}.

\bibitem{cendl}
{G. Zhigang} et~al.
\newblock \emph{CENDL-3.2: The new version of Chinese general purpose evaluated
  nuclear data library}.
\newblock EPJ Web Conf., \textbf{239}:09001, 2020.
\newblock
  \href{http://dx.doi.org/10.1051/epjconf/202023909001}{\doi{10.1051/epjconf/202023909001}}.

\bibitem{srim}
J.~F. Ziegler, M.~Ziegler and J.~Biersack.
\newblock \emph{SRIM – The stopping and range of ions in matter (2010)}.
\newblock Nuclear Instruments and Methods in Physics Research Section B: Beam
  Interactions with Materials and Atoms, \textbf{268}(11):1818--1823, 2010.
\newblock ISSN 0168-583X.
\newblock
  \href{http://dx.doi.org/https://doi.org/10.1016/j.nimb.2010.02.091}{\doi{https://doi.org/10.1016/j.nimb.2010.02.091}},
  19th International Conference on Ion Beam Analysis.

\bibitem{li2022}
L.~Li.
\newblock A Measurement of The Response of A High Purity Germanium Detector to
  Low-Energy Nuclear Recoils.
\newblock Phd dissertation, Department of Physics Duke University, 2022.
\newblock From
  \href{https://hdl.handle.net/10161/25153}{https://hdl.handle.net/10161/25153}.

\bibitem{geant4}
S.~Agostinelli et~al.
\newblock \emph{Geant4—a simulation toolkit}.
\newblock Nucl. Instrum. Meth. A, \textbf{506}(3):250--303, 2003.
\newblock ISSN 0168-9002.
\newblock
  \href{http://dx.doi.org/https://doi.org/10.1016/S0168-9002(03)01368-8}{\doi{https://doi.org/10.1016/S0168-9002(03)01368-8}}.

\bibitem{collar_comment}
J.~I. Collar and C.~M. Lewis.
\newblock \emph{Comments on "Direct measurement of the ionization quenching
  factor of nuclear recoils in germanium in the keV energy range"}, 2022.
\newblock
  \href{http://dx.doi.org/10.48550/arxiv.2203.00750}{\doi{10.48550/arxiv.2203.00750}}.

\end{thebibliography}


\begin{table*}
	\begin{subtable}[h]{\textwidth}
	\centering
    \begin{tabular*}{0.82\textwidth}{l r r r r r r r r r r r r}
    point & t (h) & $\theta$ & $\delta\theta$ & E$_{nr}$ & $\delta E_{nr}$ & $\sigma_{nr}$ & $\chi^2$/ndf & p & $\phi_{norm}$ & $q$ & $\delta q_{stat}$ & $\delta q$ \\
    \hline
6a\_6 & 2.50 & 30.6 & 0.8 & 0.97 & 0.05 & 0.15 & 1/5 & 0.96 & -0.06 & 0.181 & 0.010 & 0.013\\
\hdashline
7b\_7 & 1.25 & 33.7 & 1.1 & 1.17 & 0.07 & 0.18 & 2/3 & 0.62 & 0.11 & 0.192 & 0.015 & 0.019\\
7b\_9 & 1.25 & 33.7 & 1.1 & 1.17 & 0.07 & 0.18 & 3/4 & 0.56 & 0.24 & 0.182 & 0.011 & 0.016\\
\hdashline
9a\_11 & 1.25 & 39.0 & 0.9 & 1.54 & 0.07 & 0.19 & 15/8 & 0.06 & 0.93 & 0.190 & 0.007 & 0.011\\
9a\_9 & 1.25 & 39.0 & 0.9 & 1.54 & 0.07 & 0.19 & 6/9 & 0.71 & 0.38 & 0.183 & 0.010 & 0.013\\
\hdashline
8a\_10 & 1.25 & 39.5 & 0.9 & 1.58 & 0.07 & 0.20 & 10/10 & 0.48 & 0.67 & 0.193 & 0.007 & 0.010\\
8a\_8 & 1.25 & 39.5 & 0.9 & 1.58 & 0.07 & 0.20 & 4/10 & 0.93 & 0.47 & 0.191 & 0.008 & 0.011\\
\hdashline
11a\_11 & 1.25 & 43.8 & 0.5 & 1.92 & 0.04 & 0.22 & 14/13 & 0.35 & -0.49 & 0.190 & 0.020 & 0.020\\
11a\_7 & 1.25 & 43.8 & 0.5 & 1.92 & 0.04 & 0.22 & 16/14 & 0.31 & 0.42 & 0.182 & 0.005 & 0.006\\
\hdashline
10a\_10 & 1.25 & 44.6 & 0.5 & 1.99 & 0.04 & 0.22 & 6/13 & 0.95 & 0.08 & 0.187 & 0.004 & 0.006\\
10a\_6 & 1.25 & 44.6 & 0.5 & 1.99 & 0.04 & 0.22 & 10/10 & 0.49 & 0.86 & 0.193 & 0.006 & 0.007
\end{tabular*}
\caption{Analysis results for the coincidence distributions of runs 3 and 4 obtained with $E_{n,0}$ = 250\,keV.}\label{tab:qf_results_250keV}
\end{subtable}
	\begin{subtable}[h]{\textwidth}
	\centering
    \begin{tabular*}{0.82\textwidth}{l r r r r r r r r r r r r}
    point & t (h) & $\theta$ & $\delta\theta$ & E$_{nr}$ & $\delta E_{nr}$ & $\sigma_{nr}$ & $\chi^2$/ndf & p & $\phi_{norm}$ & $q$ & $\delta q_{stat}$ & $\delta q$ \\
    \hline
3a\_3 & 1.50 & 16.9 & 1.2 & 0.79 & 0.11 & 0.24 & 7/6 & 0.37 & -0.32 & 0.171 & 0.008 & 0.026\\
2a\_2 & 1.50 & 17.1 & 1.2 & 0.80 & 0.11 & 0.24 & 55/6 & 0.00 & -1.39 & 0.178 & 0.008 & 0.027\\
1a\_1 & 1.50 & 17.0 & 1.4 & 0.81 & 0.12 & 0.26 & 19/6 & 0.00 & -0.47 & 0.148 & 0.012 & 0.027\\
\hdashline
2b\_2 & 1.50 & 19.5 & 1.3 & 1.05 & 0.13 & 0.28 & 7/6 & 0.31 & 1.00 & 0.175 & 0.002 & 0.023\\
3b\_3 & 1.50 & 19.6 & 1.3 & 1.06 & 0.13 & 0.29 & 25/6 & 0.00 & -0.10 & 0.187 & 0.005 & 0.025\\
1b\_1 & 1.50 & 19.9 & 1.4 & 1.09 & 0.15 & 0.30 & 7/6 & 0.30 & -1.32 & 0.165 & 0.000 & 0.024\\
\hdashline
4a\_4 & 3.00 & 24.3 & 1.1 & 1.60 & 0.13 & 0.32 & 44/17 & 0.00 & 1.15 & 0.172 & 0.001 & 0.015\\
5a\_5 & 3.00 & 24.6 & 1.1 & 1.64 & 0.13 & 0.33 & 53/17 & 0.00 & 1.21 & 0.164 & 0.002 & 0.014\\
\hdashline
6a\_6 & 1.50 & 30.5 & 0.8 & 2.49 & 0.11 & 0.38 & 61/36 & 0.01 & 0.41 & 0.185 & 0.001 & 0.009\\
6a\_8 & 1.50 & 30.5 & 0.8 & 2.49 & 0.11 & 0.38 & 39/36 & 0.33 & 0.12 & 0.189 & 0.001 & 0.009\\
\hdashline
7b\_7 & 1.50 & 33.7 & 1.1 & 3.02 & 0.18 & 0.44 & 38/35 & 0.35 & 0.74 & 0.194 & 0.001 & 0.012\\
7b\_9 & 1.50 & 33.7 & 1.1 & 3.02 & 0.18 & 0.44 & 29/34 & 0.69 & 0.31 & 0.196 & 0.001 & 0.012\\
\hdashline
9a\_11 & 1.50 & 39.0 & 0.9 & 3.99 & 0.17 & 0.47 & 47/36 & 0.10 & -0.34 & 0.203 & 0.001 & 0.009\\
9a\_9 & 1.50 & 39.0 & 0.9 & 3.99 & 0.17 & 0.47 & 58/38 & 0.02 & 0.59 & 0.206 & 0.001 & 0.009\\
\hdashline
8a\_10 & 1.50 & 39.5 & 0.9 & 4.08 & 0.17 & 0.47 & 57/37 & 0.02 & -0.03 & 0.205 & 0.001 & 0.009\\
8a\_8 & 1.50 & 39.5 & 0.9 & 4.08 & 0.17 & 0.47 & 44/38 & 0.23 & -0.03 & 0.207 & 0.001 & 0.009\\
\hdashline
11a\_11 & 1.50 & 43.8 & 0.5 & 4.97 & 0.11 & 0.51 & 49/38 & 0.12 & 0.46 & 0.218 & 0.001 & 0.005\\
11a\_7 & 1.50 & 43.8 & 0.5 & 4.97 & 0.11 & 0.51 & 57/37 & 0.02 & -0.29 & 0.219 & 0.001 & 0.005\\
\hdashline
10a\_10 & 1.50 & 44.6 & 0.5 & 5.15 & 0.10 & 0.52 & 54/34 & 0.02 & 0.79 & 0.215 & 0.001 & 0.005\\
10a\_6 & 1.50 & 44.6 & 0.5 & 5.15 & 0.10 & 0.52 & 28/36 & 0.83 & -0.57 & 0.216 & 0.001 & 0.005
\end{tabular*}
\caption{Analysis results for the coincidence distributions of runs 5 and 6 obtained with $E_{n,0}$ = 640\,keV.}\label{tab:qf_results_640keV}
\end{subtable}
\caption{Analysis results for the runs 3, 4 (a) and 5, 6 (b). For each point, the irradiation time t is indicated and the selected angle $\theta$ is given with its geometrical uncertainty $\delta_{\theta}$. The corresponding selected mean nuclear recoil energy $E_{nr}$, its uncertainty $\delta E_{nr}$ and its width $\sigma_{nr}$ are obtained from a MC simulation of the setup. The $\chi^2$/ndf and p-values of the fit are reported along with the deviation $\Phi_{norm}$ in sigma units of the normalization nuisance parameter, i.e. $\Phi_{norm} = (n_{i, BF}^d-n_0^d)/\Phi_d$ following the notations of Eq. (\ref{eq:chi2_pull}). For each distribution, the best-fit of the energy independent quenching factor is given by $q$. Its statistical uncertainty is denoted by $\delta q_{stat}$, whereas $\delta q$ includes the geometrical uncertainty. The dashed lines indicate how the points were combined and shown on Fig. \ref{fig:result}.}\label{tab:results_runs3456}
\end{table*}

\begin{table*}
	\begin{subtable}[h]{\textwidth}
	\centering
    \begin{tabular*}{0.82\textwidth}{l r r r r r r r r r r r r}
    point & t (h) & $\theta$ & $\delta\theta$ & E$_{nr}$ & $\delta E_{nr}$ & $\sigma_{nr}$ & $\chi^2$/ndf & p & $\phi_{norm}$ & $q$ & $\delta q_{stat}$ & $\delta q$ \\
    \hline
3a\_3 & 3.50 & 16.8 & 1.2 & 0.60 & 0.08 & 0.19 & 11/6 & 0.10 & -0.34 & 0.159 & 0.021 & 0.031\\
2a\_2 & 3.50 & 17.0 & 1.3 & 0.61 & 0.09 & 0.19 & 17/6 & 0.01 & -0.50 & 0.166 & 0.006 & 0.024\\
1a\_1 & 3.50 & 17.1 & 1.4 & 0.61 & 0.10 & 0.20 & 6/6 & 0.38 & -0.12 & 0.159 & 0.013 & 0.028\\
\hdashline
2b\_2 & 2.00 & 19.5 & 1.3 & 0.80 & 0.10 & 0.23 & 9/6 & 0.16 & 0.73 & 0.182 & 0.015 & 0.028\\
3b\_3 & 2.00 & 19.6 & 1.3 & 0.80 & 0.10 & 0.23 & 14/6 & 0.03 & -0.52 & 0.177 & 0.007 & 0.024\\
1b\_1 & 2.00 & 19.8 & 1.4 & 0.82 & 0.12 & 0.24 & 26/6 & 0.00 & -1.16 & 0.159 & 0.011 & 0.025\\
\hdashline
4a\_4 & 5.50 & 24.3 & 1.1 & 1.22 & 0.10 & 0.26 & 12/6 & 0.07 & -0.20 & 0.175 & 0.002 & 0.015\\
5a\_5 & 5.50 & 24.6 & 1.1 & 1.25 & 0.11 & 0.26 & 15/6 & 0.02 & -0.08 & 0.167 & 0.001 & 0.014\\
\hdashline
7a\_7 & 2.00 & 29.1 & 0.8 & 1.73 & 0.09 & 0.29 & 31/18 & 0.03 & 0.72 & 0.187 & 0.002 & 0.010\\
7a\_9 & 2.00 & 29.1 & 0.8 & 1.73 & 0.09 & 0.29 & 19/17 & 0.32 & 0.48 & 0.190 & 0.002 & 0.010\\
\hdashline
6a\_8 & 2.00 & 30.6 & 0.7 & 1.90 & 0.09 & 0.30 & 32/18 & 0.02 & 0.96 & 0.186 & 0.002 & 0.009\\
\hdashline
7b\_9 & 2.00 & 33.7 & 1.1 & 2.30 & 0.14 & 0.35 & 29/31 & 0.55 & 0.51 & 0.195 & 0.002 & 0.012\\
6b\_6 & 2.00 & 34.1 & 1.0 & 2.35 & 0.13 & 0.35 & 31/32 & 0.54 & -0.11 & 0.202 & 0.001 & 0.012\\
6b\_8 & 2.00 & 34.1 & 1.0 & 2.35 & 0.13 & 0.35 & 26/34 & 0.85 & 0.43 & 0.203 & 0.002 & 0.012\\
\hdashline
9a\_11 & 3.50 & 39.0 & 0.9 & 3.04 & 0.13 & 0.38 & 42/37 & 0.27 & 0.29 & 0.199 & 0.001 & 0.009\\
9a\_9 & 2.00 & 39.0 & 0.9 & 3.04 & 0.13 & 0.38 & 47/33 & 0.05 & -0.66 & 0.200 & 0.001 & 0.009\\
8a\_10 & 3.50 & 39.5 & 0.9 & 3.11 & 0.13 & 0.38 & 47/37 & 0.12 & -0.36 & 0.204 & 0.001 & 0.009\\
8a\_8 & 2.00 & 39.5 & 0.9 & 3.11 & 0.13 & 0.38 & 48/33 & 0.04 & -0.46 & 0.202 & 0.001 & 0.009\\
\hdashline
11a\_11 & 2.00 & 43.8 & 0.5 & 3.78 & 0.08 & 0.41 & 29/33 & 0.66 & -0.44 & 0.216 & 0.001 & 0.005\\
11a\_7 & 3.50 & 43.8 & 0.5 & 3.78 & 0.08 & 0.41 & 42/37 & 0.28 & 0.13 & 0.212 & 0.001 & 0.005\\
\hdashline
10a\_10 & 2.00 & 44.6 & 0.5 & 3.92 & 0.09 & 0.41 & 34/36 & 0.57 & -0.09 & 0.214 & 0.001 & 0.005\\
10a\_6 & 3.50 & 44.6 & 0.5 & 3.92 & 0.09 & 0.41 & 46/37 & 0.16 & -0.21 & 0.215 & 0.001 & 0.005
\end{tabular*}
\caption{Analysis results for the coincidence distributions of runs 7, 8, 9 obtained with $E_{n,0}$ = 500\,keV.}\label{tab:qf_results_500keV}
\end{subtable}
	\begin{subtable}[h]{\textwidth}
	\centering
    \begin{tabular*}{0.82\textwidth}{l r r r r r r r r r r r r}
    point & t (h) & $\theta$ & $\delta\theta$ & E$_{nr}$ & $\delta E_{nr}$ & $\sigma_{nr}$ & $\chi^2$/ndf & p & $\phi_{norm}$ & $q$ & $\delta q_{stat}$ & $\delta q$ \\
    \hline
3a\_3 & 1.00 & 16.8 & 1.2 & 0.96 & 0.14 & 0.30 & 10/6 & 0.15 & -0.05 & 0.181 & 0.010 & 0.027\\
2a\_2 & 1.00 & 17.0 & 1.2 & 0.98 & 0.14 & 0.31 & 2/6 & 0.91 & -0.11 & 0.195 & 0.002 & 0.027\\
1a\_1 & 1.00 & 17.1 & 1.4 & 0.99 & 0.16 & 0.33 & 12/6 & 0.06 & -0.15 & 0.142 & 0.001 & 0.023\\
\hdashline
2b\_2 & 1.30 & 19.5 & 1.3 & 1.29 & 0.16 & 0.36 & 4/6 & 0.67 & 0.63 & 0.192 & 0.005 & 0.025\\
3b\_3 & 1.30 & 19.6 & 1.3 & 1.30 & 0.17 & 0.36 & 2/6 & 0.90 & -0.02 & 0.199 & 0.004 & 0.025\\
1b\_1 & 1.30 & 19.8 & 1.4 & 1.33 & 0.18 & 0.39 & 16/6 & 0.01 & 0.02 & 0.163 & 0.006 & 0.023\\
\hdashline
4a\_4 & 2.30 & 24.4 & 1.1 & 1.97 & 0.17 & 0.42 & 30/17 & 0.03 & 0.47 & 0.180 & 0.002 & 0.015\\
5a\_5 & 2.30 & 24.6 & 1.1 & 2.01 & 0.17 & 0.42 & 18/27 & 0.90 & 0.92 & 0.164 & 0.002 & 0.014\\
\hdashline
7a\_7 & 1.30 & 29.1 & 0.8 & 2.79 & 0.14 & 0.47 & 30/29 & 0.40 & 0.36 & 0.192 & 0.003 & 0.010\\
7a\_9 & 1.00 & 29.1 & 0.8 & 2.79 & 0.14 & 0.47 & 15/20 & 0.75 & 0.95 & 0.198 & 0.003 & 0.010\\
\hdashline
6b\_6 & 1.30 & 34.1 & 1.0 & 3.79 & 0.22 & 0.56 & 17/28 & 0.94 & 0.50 & 0.221 & 0.002 & 0.013\\
6b\_8 & 1.00 & 34.1 & 1.0 & 3.79 & 0.22 & 0.56 & 26/24 & 0.35 & -0.73 & 0.220 & 0.002 & 0.013\\
\hdashline
9a\_11 & 1.00 & 39.0 & 0.9 & 4.89 & 0.21 & 0.60 & 16/26 & 0.95 & 0.25 & 0.211 & 0.003 & 0.009\\
9a\_9 & 1.30 & 39.0 & 0.9 & 4.89 & 0.21 & 0.60 & 20/25 & 0.76 & 0.27 & 0.215 & 0.002 & 0.009\\
\hdashline
8a\_10 & 1.00 & 39.5 & 0.9 & 5.01 & 0.21 & 0.60 & 28/31 & 0.63 & -0.30 & 0.223 & 0.002 & 0.010\\
8a\_8 & 1.30 & 39.5 & 0.9 & 5.01 & 0.21 & 0.60 & 17/28 & 0.95 & 0.34 & 0.222 & 0.002 & 0.010\\
\hdashline
11a\_11 & 1.30 & 43.8 & 0.5 & 6.10 & 0.14 & 0.66 & 18/29 & 0.95 & 0.36 & 0.224 & 0.002 & 0.005\\
11a\_7 & 1.00 & 43.8 & 0.5 & 6.10 & 0.14 & 0.66 & 33/25 & 0.12 & 0.10 & 0.227 & 0.002 & 0.006\\
\hdashline
10a\_10 & 1.30 & 44.6 & 0.5 & 6.32 & 0.14 & 0.65 & 28/29 & 0.50 & -0.28 & 0.236 & 0.002 & 0.006\\
10a\_6 & 1.00 & 44.6 & 0.5 & 6.32 & 0.14 & 0.65 & 29/27 & 0.34 & 0.39 & 0.233 & 0.002 & 0.006
\end{tabular*}
\caption{Analysis results for the coincidence distributions of runs 10 and 11 obtained with $E_{n,0}$ = 800\,keV.}\label{tab:qf_results_800keV}
\end{subtable}
\caption{Analysis results for the runs 7, 8, 9 (a) and 10, 11 (b). The description of the columns is identical to the one in Tab. \ref{tab:results_runs3456}.}\label{tab:results_runs7891011}
\end{table*}

\end{document}